\documentclass[12pt]{article}
\pdfoutput=1

\DeclareFontFamily{OT1}{pzc}{}
\DeclareFontShape{OT1}{pzc}{m}{it}{<-> s * [1.10] pzcmi7t}{}
\DeclareMathAlphabet{\mathpzc}{OT1}{pzc}{m}{it}

\newcommand{\bb}[1]{\mathbb{#1}}
\newcommand{\wt}[1]{\widetilde{#1}}
\newcommand{\ol}[1]{\overline{#1}}
\newcommand{\mc}[1]{\mathcal{#1}}

\usepackage{stmaryrd}

\usepackage{draft} 
\usepackage[weather]{ifsym}

\usepackage{hyperref}
\usepackage{graphicx,color,subfig}
\usepackage{cite}
\usepackage{mciteplus}
\usepackage{skak}
\usepackage{empheq}
\usepackage{tikz}
\usepackage{bbm}

\DeclareFontFamily{OT1}{pzc}{}
\DeclareFontShape{OT1}{pzc}{m}{it}{<-> s * [1.10] pzcmi7t}{}
\DeclareMathAlphabet{\mathpzc}{OT1}{pzc}{m}{it}

\usetikzlibrary{calc}
\usetikzlibrary{snakes}
\usetikzlibrary{arrows.meta}
\usetikzlibrary{decorations.pathmorphing}
\usetikzlibrary{decorations.markings}
\usetikzlibrary{bending}
\tikzset{snake it/.style={decorate, decoration=snake}}
\usetikzlibrary{shapes.misc}
\tikzset{cross/.style={cross out, draw=black, minimum size=2*(#1-\pgflinewidth), inner sep=0pt, outer sep=0pt},
cross/.default={1pt}}

\usepackage[T1]{fontenc}
\usepackage{esint}
\usepackage{lmodern}

\def\be#1\ee{\begin{align}#1\end{align}}

\definecolor{dark green}{rgb}{0.7,1,0.64}

\usepackage{listings}
\usepackage{xcolor}

\definecolor{codegreen}{rgb}{0,0.6,0}
\definecolor{codegray}{rgb}{0.5,0.5,0.5}
\definecolor{codepurple}{rgb}{0.58,0,0.82}
\definecolor{backcolour}{rgb}{0.95,0.95,0.92}

\lstdefinestyle{myStyle}{
    belowcaptionskip=1\baselineskip,
    breaklines=true,
    frame=none,
    numbers=none, 
    basicstyle=\footnotesize\ttfamily,
    keywordstyle=\bfseries\color{green!40!black},
    commentstyle=\itshape\color{purple!40!black},
    identifierstyle=\color{blue},
    backgroundcolor=\color{gray!10!white},
    tabsize=2,
}

\usepackage{array}

\begin{document}

\unitlength = .8mm

\begin{titlepage}

\begin{center}

\hfill \\
\hfill \\
\vskip 1cm

\title{Conformal Perturbation Theory and Tachyon-Dilaton Eschatology via String Fields}

\author{Ben Mazel, Joshua Sandor, Charles Wang, Xi Yin}

\address{Jefferson Physical Laboratory, Harvard University, 
Cambridge, MA 02138 USA
}

\email{bmazel@g.harvard.edu, jsandor@fas.harvard.edu, charles\_wang@g.harvard.edu, xiyin@fas.harvard.edu}

\end{center}

\abstract{We analyze deformations of two-dimensional conformal field theory (CFT) from the perspective of classical bosonic closed string field theory (SFT). The latter can be viewed as a version of Wilsonian renormalization group (RG) improved conformal perturbation theory, where the renormalization scheme is defined through the choice of string vertices in the construction of SFT. Furthermore, the CFT data at the RG fixed point can be recovered from the spectrum and amplitudes of string field fluctuations. As applications, we construct the Horowitz-Polchinski ``string star'' solution in SFT, and a solution of tachyon-dilaton condensation that deforms the noncompact boson to minimal models by creating a pair of ``Runkel-Watts walls''.
}

\vfill

\end{titlepage}

\eject

\begingroup
\hypersetup{linkcolor=black}

\tableofcontents

\endgroup

\section{Introduction}

Two-dimensional conformal field theories (CFTs) play an essential role in understanding critical phenomena \cite{Zinn-Justin:2002ecy}, and separately, in formulating string theory from the worldsheet perspective \cite{Polchinski:1998rq}. Despite a wealth of solvable examples \cite{Belavin:1984vu, Gepner:1986wi, Candelas:1990rm, Zamolodchikov:1995aa, DiFrancesco:1997nk, Teschner:1997ft}, it is suspected that much of the landscape of 2D unitary CFTs remains unexplored \cite{Gukov:2002nw, Hellerman_2011, Collier:2016cls, Antunes:2022vtb}. A basic tool for analyzing the connection between different CFTs, through either marginal deformations or renormalization group (RG) flows, is conformal perturbation theory \cite{Zamolodchikov:1987ti}. In practice, beyond leading order analysis for small deformations, conformal perturbation theory is extremely unwieldy. Much of the difficulty lies in the implementation of a consistent regularization and renormalization scheme (e.g. via point-splitting \cite{Gaberdiel:2008fn}), where one must wrestle with complicated scheme-dependent artifacts before arriving at unambiguously defined physical observables.

From the string theory standpoint, 2D CFTs are recipes of the worldsheet theory that defines consistent spacetime backgrounds in which the strings propagate and interact, at least at the perturbative level. In the on-shell formulation of critical bosonic string theory, in particular, consistent spacetime backgrounds are described by a ``matter'' CFT of central charge $c=26$, which is combined with the conformal $bc$ ghost system to produce the full worldsheet CFT that admits a nilpotent BRST charge and a BRST-exact stress-energy tensor. A deformation of the matter CFT, at least in the case where the central charge is preserved, amounts to a deformation of the spacetime background.

%\footnote{At the loop level, there may be infrared divergences, particularly due to the propagation of tachyons, but this is not relevant at the level of classical SFT considered in this paper.}

On the other hand, one may equivalently interpret a deformation of the spacetime background in string theory as a coherent state of string excitations in the original, undeformed, background. This can be precisely formulated in the framework of string field theory (SFT), in which the string excitations are represented through spacetime fields \cite{Witten:1985cc, Zwiebach:1992ie}. In particular, the string scattering amplitudes, which in the on-shell formalism are constructed by integrating suitable correlators of the worldsheet CFT over the moduli space of punctured Riemann surfaces \cite{Polchinski:1998rq}, are reproduced in SFT by summing over Feynman diagrams that arise from an action functional.

For any given CFT $\mathbb{A}$ of interest, whose central charge is denoted $c_{\mathbb A}$, we can construct a classical bosonic closed SFT whose worldsheet matter CFT consists of $\mathbb{A}$ and an auxiliary CFT of central charge $26-c_\mathbb{A}$, and seek solutions to the string field equation of motion that represents deformations of $\mathbb{A}$ (but not the auxiliary CFT). In particular, corresponding to a family of CFTs $\mathbb{A}_\lambda$ obtained by deforming $\mathbb{A}$ with an exactly marginal operator ${\cal O}$, we can construct a family of SFT solutions $\Psi_\lambda$ of the form
\ie\label{psigenform}
\Psi_\lambda = \sum_{n=1}^\infty \lambda^n \Psi_n \in {\cal H}_{\mathbb{A}} \otimes {\cal V}_0\otimes {\cal H}_{bc},~~~~{\rm with}~~\Psi_1 = c\widetilde c {\cal O},
\fe
where $\lambda$ is the deformation parameter, ${\cal H}_{\mathbb{A}}$ is the Hilbert space of the (undeformed) CFT $\mathbb{A}$, ${\cal V}_0$ is the identity Virasoro module of the auxiliary CFT, and ${\cal H}_{bc}$ is the state space of the $bc$ ghost system. The ``background independence'' of SFT \cite{Sen:1990hh, Sen:1990na, Sen:1992pw, Sen:1993mh, Sen:1993kb} is such that the spectral data of the deformed CFT $\mathbb{A}_\lambda$ can be recovered in the SFT formalism from the spectrum and scattering amplitudes of string field fluctuations around $\Psi_\lambda$.

An important subtlety arises for noncompact CFT $\mathbb{A}$, whose exactly marginal deformation may change the central charge. In this case, we would find a cohomological obstruction to solving the string field equation, indicating that the shift of central charge of the worldsheet CFT(away from criticality) would result in an inconsistent string background. This problem can be circumvented if we replaced $\mathbb{A}$ by its tensor product with a linear dilaton CFT $\mathbb{D}_\beta$, and consider simultaneous deformation of $\mathbb{A}$ and of the linear dilaton background charge $\beta$, so that the total central charge remains invariant. We will see that including the linear dilaton sector allows for removing the cohomological obstruction, and we can construct a SFT solution
\ie\label{psilambdagen}
\Psi_\lambda  =\sum_{n=1}^\infty \lambda^n \Psi_n \in {\cal H}_{\mathbb{A}} \otimes {\cal H}_{\mathbb{D}} \otimes {\cal V}_0 \otimes {\cal H}_{bc}
\fe
that corresponds to the deformed CFT $\mathbb{A}_\lambda\otimes \mathbb{D}_{\beta(\lambda)}$ where $\beta(\lambda)$ is the deformed background charge.

A similar construction is applicable to a short RG flow generated by a slightly relevant deformation of $\mathbb{A}$. In this case, one must balance terms of different orders in the string field equation; consequently, the expansion parameter $\lambda$ in (\ref{psilambdagen}) will be fixed by the small parameter that the characterizes the short RG flow \cite{Mukherji:1991tb}. Indeed, $\lambda$ can be viewed as a renormalized coupling, and $\Psi_\lambda$ itself may be viewed as a sort of Wilsonian effective action at the RG fixed point. Here, a choice of renormalization scheme of the Wilsonian RG is implicitly specified in the SFT formalism through the choice of string vertices; as such, scheme dependence can be absorbed into string field redefinition and will drop out of physical observables \cite{Hata:1993gf}. This construction also describes closed string tachyon condensation \cite{Headrick:2004hz} in a perturbatively controlled setting.\footnote{See \cite{Scheinpflug:2023lfn} for a recent investigation.}

The $bc$ ghost system in the SFT construction may seem out of place from the conformal perturbation theory standpoint. In fact, we will see that the ghosts play an essential role in treating the dilaton deformation in the target space, which is awkward to handle in the conventional language of conformal perturbation theory. We will also exploit the flexibility of gauge choice of SFT, and the flexibility in the choice of string vertices or the string field frame. In particular, we will make extensive use of a ``flat-vertex'' frame, related to the conventional closed SFT construction \cite{Zwiebach:1992ie} by a field redefinition, in which the string bracket used in writing the nonlinear terms in the SFT equation can be computed simply without the need for nontrivial conformal transformations on the string fields, that often leads to dramatic technical simplifications.

A simple application of our approach is the construction of the Horowitz-Polchinski ``string star'' solution \cite{Horowitz:1996nw, Horowitz:1997jc}, that describes a self-gravitating thermal ensemble of strings, in the SFT framework for slow-varying fields. This construction amounts to an RG-improved conformal perturbation theory treatment of the corresponding worldsheet CFT deformation, and recovers the known results from the spacetime massless effective field theory.

In a second and technically more involved application, we study a perturbatively-marginal deformation of the noncompact free boson generated by the operator $\cos(\sqrt{2} X)$ at the leading order. The second order SFT solution reveals a negative shift of the central charge. Further analysis of the SFT solution in the slow-varying field approximation in the large field regime reveals a spatially-modulated tachyon and dilaton profile in the target space. Surprisingly, the SFT solution reveals a finite space in between a pair of walls, each wall being identified with the Runkel-Watts theory on the worldsheet. The standing wave of the tachyon between the walls reproduce the low-lying spectrum of the $A_{k}$ minimal model at large $k$, including the detailed quantization property. This leads us to conclude that the result of the $\cos(\sqrt{2} X)$ deformation is in fact none other than the $A_{k}$ minimal model, but described by a rather nontrivial string background of ``tachyon-dilaton eschatology''.

The rest of the paper is organized as follows. In section \ref{sec:sftreview} we review the recipes that define a classical bosonic closed string field theory, including the worldsheet CFT used to define the space of string fields, the construction of string vertices and related string field brackets, and the SFT equation of motion. We also discuss deformation of string vertices and the construction of a ``flat-vertex'' string field frame that will be useful for subsequent analysis of SFT solutions.
In section \ref{sec:pertsol}, we present a framework for analyzing CFT deformations using SFT. Section \ref{sec:pertrg} analyzes short RG flows as perturbative SFT solutions; the results of this section are already known in \cite{Mukherji:1991tb}, but we include a derivation in the flat-vertex string field frame for completeness. In section \ref{sec:recover} we describe the strategy for recovering deformed CFT data from the spectrum and on-shell amplitudes of fluctuations around a SFT solution. Section \ref{sec:hp} presents the SFT solution that describes the Horowitz-Polchinski string star in the slow-varying field approximation. In section \ref{sec:escha} we study the $\cos(\sqrt{2} X)$ deformation of the noncompact boson by its corresponding SFT ``tachyon-dilaton eschatology'' solution, which unveils the emergence of Runkel-Watts walls and the minimal model spectrum from standing tachyon waves. Some concluding remarks are given in section \ref{sec:discuss}. Useful technical details concerning the CFT data of Runkel-Watts theory are given in Appendix \ref{sec:runkelwatts}.

\section{Classical string field theory}
\label{sec:sftreview}

\subsection{The worldsheet CFT and the space of string fields}

The worldsheet CFT consists of a $c = 26$ matter CFT and the $bc$ ghost system \cite{Polchinski:1998rq}. The latter involves anticommuting holomorphic fields $b(z)$ and $c(z)$, of conformal weight $2$ and $-1$ respectively, that obey the OPE
\ie
b(z) c(0) = {1\over z} + \sum_{n=0}^\infty {z^n\over n!} :(\partial^n b) c(0):, 
\fe
and their anti-holomorphic counterparts $\wt b(\bar z)$ and $\wt c(\bar z)$. The normal ordering symbol will henceforth be omitted for operators defined as product of fields at coinciding points. The stress-energy tensor of the $bc$ ghost system is
\ie
T^{\rm gh} = - (\partial b) c - 2 b \partial c,
\fe
with central charge $-26$.

Importantly, the worldsheet CFT admits a BRST symmetry, generated by the charge
\ie
Q_B = \oint \frac{dz}{2 \pi i} j_B(z)+ \oint \frac{d \overline z}{2 \pi (- i)} \wt j_B(\ol z),
\fe
where the holomorphic component of the BRST current $j_B(z)$ is given by
\ie\label{jbrst}
j_B = c T^{\rm m} + b c \partial c + \frac{3}{2}\partial^2 c .
\fe
Here $T^{\rm m}$ is the $c=26$ matter stress-energy tensor. The last total derivative term on the RHS of (\ref{jbrst}) is unimportant for the definition of $Q_B$, but is included to make $j_B$ a Virasoro primary. The key properties of the BRST charge $Q_B$ are its nilpotency, namely $Q_B^2 = 0$, and
\ie
Q_B\cdot b = T \equiv T^{\rm m} + T^{\rm gh}.
\fe
We denote by ${\cal H}^{\rm m}$ the Hilbert space of the matter CFT, in the sense of the space of states on $S^1$ or equivalently the space of local operators, and likewise by ${\cal H}^{\rm gh}$ the space of states of the $bc$ ghost system. In particular, the lowest weight subspace of ${\cal H}^{\rm gh}$ is spanned by $c\widetilde c$, $c\partial c \widetilde c$, $c\wt c \bar\partial\wt c$, and $c\partial c \wt c \bar\partial\wt c$. The full worldsheet CFT state space is ${\cal H}^{\rm m}\otimes {\cal H}^{\rm gh}$.

As usual we write $b_n = \oint {dz\over 2\pi i} z^{n+1} b(z)$, $c_n =\oint {dz\over 2\pi i} z^{n-2} c(z)$, $L_n = \oint {dz\over 2\pi i} z^{n+1} T(z)$, and similarly for their anti-holomorphic counterparts. It will be convenient to define
\ie
b_0^{\pm} \equiv b_0 \pm \wt b_0, ~~~~c_0^{\pm} \equiv \frac{1}{2} (c_0 \pm \wt c_0), ~~~~ L_0^{\pm} \equiv L_0 \pm \wt L_0.
\fe
We further define $\bb P^-$ to be the projector onto the subspace annihilated by $L_0^-$, and $\bb P^+$ to be the projector onto the subspace annihilated by $L_0^+$, or more generally the subspace on which $L_0^+$ acts nilpotently. Note that ${\cal H}^{\rm gh}$ is equipped with the BPZ inner product, which is not positive definite; nonetheless, one typically works with subspaces of ${\cal H}^{\rm gh}$ with a definite ghost number (assigning $+1$ to $c$ and $-1$ to $b$), where projectors can be defined analogously to those of a Hilbert space.

The space of string fields, denoted $\hat {\cal H}$, is a subspace of the CFT state space defined as \cite{Zwiebach:1992ie}
\ie\label{hsft}
\hat{\cal H} = \{\Psi \in {\cal H}^{\rm m}\otimes {\cal H}^{\rm gh}: ~ b_0^- \Psi = L_0^- \Psi=0 \}.
\fe

\subsection{String vertices and string field brackets}

\begin{figure}[h!]
	\def\svgwidth{1\linewidth}
	\centering{
		%% Creator: Inkscape 1.1.2 (b8e25be8, 2022-02-05), www.inkscape.org
%% Accompanies image file 'cutglue.pdf' (pdf, eps, ps)
\begingroup%
  \makeatletter%
  \providecommand\color[2][]{%
    \errmessage{(Inkscape) Color is used for the text in Inkscape, but the package 'color.sty' is not loaded}%
    \renewcommand\color[2][]{}%
  }%
  \providecommand\transparent[1]{%
    \errmessage{(Inkscape) Transparency is used (non-zero) for the text in Inkscape, but the package 'transparent.sty' is not loaded}%
    \renewcommand\transparent[1]{}%
  }%
  \providecommand\rotatebox[2]{#2}%
  \newcommand*\fsize{\dimexpr\f@size pt\relax}%
  \newcommand*\lineheight[1]{\fontsize{\fsize}{#1\fsize}\selectfont}%
  \ifx\svgwidth\undefined%
    \setlength{\unitlength}{1153.7007874bp}%
    \ifx\svgscale\undefined%
      \relax%
    \else%
      \setlength{\unitlength}{\unitlength * \real{\svgscale}}%
    \fi%
  \else%
    \setlength{\unitlength}{\svgwidth}%
  \fi%
  \global\let\svgwidth\undefined%
  \global\let\svgscale\undefined%
  \makeatother%
  \begin{picture}(1,0.22113022)%
    \lineheight{1}%
    \setlength\tabcolsep{0pt}%
    \put(0,0){\includegraphics[width=\unitlength,page=1]{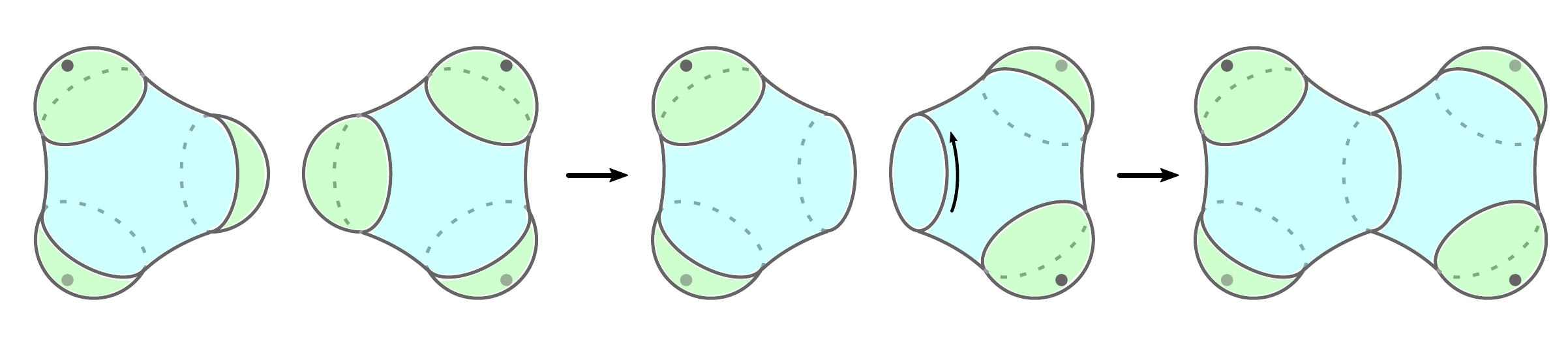}}%
    \put(0.02493452,0.19124352){\makebox(0,0)[lt]{\lineheight{1.25}\smash{\begin{tabular}[t]{l}1\end{tabular}}}}%
    \put(0.02641977,0.01170889){\makebox(0,0)[lt]{\lineheight{1.25}\smash{\begin{tabular}[t]{l}2\end{tabular}}}}%
    \put(0.32545584,0.19091369){\makebox(0,0)[lt]{\lineheight{1.25}\smash{\begin{tabular}[t]{l}3\end{tabular}}}}%
    \put(0.32659972,0.01210174){\makebox(0,0)[lt]{\lineheight{1.25}\smash{\begin{tabular}[t]{l}4\end{tabular}}}}%
    \put(0.61765501,0.10290059){\makebox(0,0)[lt]{\lineheight{1.25}\smash{\begin{tabular}[t]{l}$\theta$\end{tabular}}}}%
  \end{picture}%
\endgroup%
	\caption{Plumbing a pair of 3-punctured spheres with a twist by angle $\theta$.
			\label{cutandglue}
	}}
\end{figure}
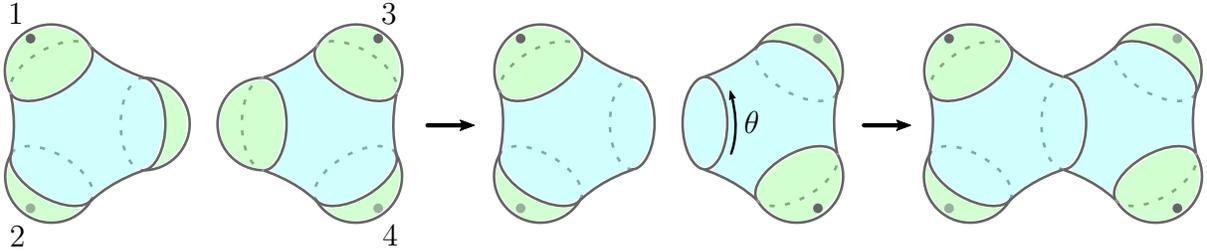

Let ${\cal P}_{0,n}$ be the space of $n$-punctured Riemann sphere, together with a set of coordinate maps
\ie\label{coordmap}
z = f_i(w_i).
\fe
from the unit $w_i$-discs to the local chart $D_i$ that contains the $i$-th puncture at $z=z_i\equiv f_i(0)$, modulo the overall ${\rm PSL}(2,\mathbb{C})$ conformal Killing group action. ${\cal P}_{0,n}$ can be viewed as an infinite-dimensional fiber bundle over the moduli space ${\cal M}_{0,n}$ of the $n$-punctured Riemann sphere. A basic operation is the plumbing map
\ie\label{plumb}
\#_{i,j}: ~{\cal P}_{0,n_1}\times {\cal P}_{0,n_2}\times S^1 \to {\cal P}_{n_1+n_2-2},
\fe
which takes an $n_1$-puncture sphere $\Sigma$ with local charts $D_1, \cdots, D_{n_1}$ containing the punctures, an $n_2$-punctured sphere $\Sigma'$ with local charts $D_1',\cdots, D_{n_2}'$ containing the punctures, and a twist angle $\theta\in [0,2\pi)$, to the $(n_1+n_2-2)$-punctured Riemann sphere formed by the joining $\Sigma\backslash D_i$ to $\Sigma'\backslash D'_j$ via the gluing map
\ie\label{gluemap}
w_i = e^{i\theta} / w'_j,
\fe
where $w_i$ and $w'_j$ are the local coordinates for $D_i$ and $D'_j$, along $|w'_j| = |w_i|=1$. This is portrayed in Figure \ref{cutandglue} for $n_1=n_2=3$.

Let $\hat {\cal P}_{0,n}$ be the quotient of ${\cal P}_{0,n}$ by the constant phase rotations of the argument of each coordinate map $f_i$ independently, which can also be viewed as a fiber bundle over ${\cal M}_{0,n}$. (\ref{plumb}) induces a plumbing map at the level of chains
\ie\label{sharpij}
\hat\#_{i,j}: ~ &C_{d_1}(\hat{\cal P}_{0,n_1})\times C_{d_2}(\hat{\cal P}_{0,n_2}) \to C_{d_1+d_2+1}(\hat{\cal P}_{n_1+n_2-2})
\\
&g_1 \times  g_2 \mapsto \#_{i,j}(g_1\times g_2\times S^1)
\fe
where by a slight abuse of notation we have denoted by $g_a$ a $d_a$-dimensional chain in $\hat{\cal P}_{0,n_a}$ as well as its representative in ${\cal P}_{0,n_a}$.

The $n$-point ``string vertex'' is a $2n-6$ dimensional chain $\Gamma_n$ in $\hat {\cal P}_{0,n}$ that is symmetric with respect to permutations on the $n$ punctures along with the coordinate maps around the punctures, and subject to the following conditions. Firstly, $\Gamma_3$ projects onto ${\cal M}_{0,3}$ itself (which is a point) with multiplicity 1, and $\Gamma_n$ (for $n\geq 4$) projects onto a top-dimensional chain in ${\cal M}_{0,n}$ that does not meet the boundary of the latter (where some of the punctures collide). Furthermore, the $\Gamma_n$'s satisfy the ``geometric master equation''
\ie\label{geometricmeq}
\partial \Gamma_n = - {1\over 2} \sum_{\ell=3}^{n-1} {n\choose \ell-1} \left[ \hat\#( \Gamma_\ell\times  \Gamma_{n+2-\ell}) \right]^{\rm Sym} ,
\fe
where $\hat\#$ is the plumbing map at the level of chains (\ref{sharpij}) with any choice of punctures $i,j$, and $[\cdots]^{\rm Sym}$ stands for the symmetrization over the $n$ punctures and coordinate maps.

As an explicit example, a possible choice of $\Gamma_3$ is the punctures at $z_1 = 0$, $z_2=1$, $z_3=-1$, with the coordinate maps
\ie\label{f123}
f_1(w_1) = {w_1\over 3r},~~~ f_2(w_2) = {1+{w_2\over 3r}\over 1-{w_2\over r}},~~~ f_3(w_3) = {-1+{w_3\over 3r}\over 1+{w_3\over r}},
\fe
where $r$ is a parameter chosen in the range $r\geq 1$. Note that $f_1, f_2, f_3$ are cyclicly permuted by the ${\rm PSL}(2,\mathbb{C})$ map $z\mapsto {1+z\over 1- 3z}$, and thus $\Gamma_3$ is symmetric. Note that when $r=1$, the discs $D_i=\{z=f_i(w):|w|\leq 1\}$ touch one another, but their union does not cover the Riemann sphere.\footnote{This is different from the 3-point string vertex defined via minimal area metric \cite{Zwiebach:1992ie}; the choice (\ref{f123}) is such that the coordinate maps are themselves ${\rm PSL}(2,\mathbb{C})$, which will be convenient for perturbative SFT computations, but ``less efficient'' in the sense that the 4-point string vertex $\Gamma_4$ will need to cover a bigger domain of the moduli space.}

The plumbing of a pair of 3-punctured spheres with the coordinate maps (\ref{f123}), by joining $\partial D_1$ with $\partial D_1'$ via (\ref{gluemap}), results in a 4-punctured sphere with punctures at $z=1, -1, {q\over 9r^2}, -{q\over 9r^2}$, where $q=e^{i\theta}$, along with four coordinate maps. In particular, the cross ratio that parameterizes this 4-punctured sphere in its moduli space ${\cal M}_{0,4}$ is $t={36 q r^2\over (q+9 r^2)^2}$. It follows from the geometric master equation that the projection of the 4-point string vertex $\Gamma_4$ onto ${\cal M}_{0,4}$ must cover (with multiplicity 1) the domain
\ie\label{gam4d}
%&\Big\{t\in \mathbb{C}: ~{|t|^{1\over 2} \over|1+\sqrt{1-t}|}, {|1-t|^{1\over 2} \over|1+\sqrt{t}|},{1 \over|\sqrt{t}+\sqrt{t-1}|} > 3r \Big\}.
\Big\{t\in \mathbb{C}: ~ {1+|1-t| \over |t|} , \, {1+|t| \over |1-t|} , \, {|t|+|1-t| }  > {1\over 2}\big(9r^2+{1\over 9r^2}\big) \Big\}.
\fe
Furthermore, at the boundary of (\ref{gam4d}), the coordinate maps on the 4-puncture sphere must agree with those obtained by the plumbing of a pair of $\Gamma_3$'s.

The $n$-vertex {\it of string fields} is a graded symmetric $n$-linear map $\{\cdot\}: \hat{\cal H}^{\otimes n}\to \mathbb{C}$ defined by\footnote{The sign in the normalization factor is in agreement with \cite{Sen:1993kb}, correcting that of \cite{Zwiebach:1992ie}.}
\ie\label{nvertex}
\{ \Psi^{\otimes n} \} = {1\over (-2\pi i)^{n-3}} \int_{\Gamma_n}  \left\langle e^{\cal B} \prod_{i=1}^n [\Psi(0)]^{f_i} \right\rangle,
\fe
where $\langle\cdots\rangle$ stands for the worldsheet CFT correlator. $[\Psi(0)]^{f_i}$ is the conformal transformation of the string field $\Psi$, viewed as an operator inserted at the origin of the $w_i$-disc, by the map (\ref{coordmap}) to an operator inserted at $z=z_i$ on the Riemann sphere. ${\cal B}$ is an operator-valued 1-form on ${\cal P}_{0,n}$ built out of the $b$ ghost, 
\ie
{\cal B} = -\sum_{i=1}^n \oint_{\partial D_i} {dz\over 2\pi i} \delta f_i(w_i) b(z) +c.c.
\fe
Here $\delta f_i$ are interpreted as 1-forms on ${\cal P}_{0,n}$. If $\Gamma_n$ takes the form of a section of $\pi:{\cal P}_{0,n}\to {\cal M}_{0,n}$ over a moduli domain, the coordinate maps may be expressed as $z=f_i(w_i; t)$, where $t=(t^1,\cdots, t^{n-3})$ are complex coordinates on ${\cal M}_{0,n}$. In this case, ${\cal B}$ may also be viewed as a 1-form on $\pi(\Gamma_n)\subset {\cal M}_{0,n}$, where the integration on the RHS of (\ref{nvertex}) is evaluated.

An essential property of the string field vertices, which follows from (\ref{geometricmeq}), is that they obey the Batalin-Vilkovisky (BV) master equation
\ie\label{bvmaster}
n \{ Q_B \Psi\otimes \Psi^{\otimes (n-1)} \} = -  {1\over 2} \sum_{\ell=2}^{n-2} {n\choose \ell} \sum_I \{ \Psi^{\otimes\ell} \otimes \phi_I  \} \{\phi^{\vee I} \otimes \Psi^{\otimes (n-\ell)} \} ,
\fe
where $\{\phi_I\}$ and $\{\phi^{\vee I}\}$ are a pair of dual bases of $\hat{\cal H}$ that obey
\ie
\langle \phi^{\vee I}| c_0^-| \phi_J\rangle = - \langle \phi_J| c_0^-| \phi^{\vee I}\rangle = \delta^I_J, ~~~~ {\bf 1}_{\hat{\cal H}} = \sum_I |\phi_I\rangle \langle\phi^{\vee I} | c_0^- = - \sum_I |\phi^{\vee I}\rangle \langle \phi_I| c_0^-.
\fe
Here $\langle\cdot|$ stands for BPZ conjugate.

The $n$-string (field) bracket is a graded symmetric $n$-linear map $[\cdot]: \hat{\cal H}^{\otimes n}\to \hat{\cal H}$, related to the $(n+1)$-vertex of string fields by 
\ie\label{bracketdef}
\langle \Phi | c_0^- |[\Psi^{\otimes n}]\rangle = \{ \Phi\otimes \Psi^{\otimes n} \}
\fe
for all $\Phi_i\in \hat{\cal H}$. By definition, $[\Psi^{\otimes n}]$ is annihilated by $b_0^-$ and $L_0^-$ (cf. (\ref{hsft})), and is consequently determined by the defining property (\ref{bracketdef}).\footnote{This can be seen by writing, for any spinless $\chi\in{\cal H}$ (i.e. annihilated by $L_0^-$, but not necessarily by $b_0^-$), $\langle\chi |[\Psi^{\otimes n}]\rangle = \langle\chi|(b_0^-c_0^- + c_0^- b_0^-) |[\Psi^{\otimes n}]\rangle = \langle\chi|b_0^- c_0^-  |[\Psi^{\otimes n}]\rangle $. Now $b_0^-\chi$ is annihilated by both $L_0^-$ and $b_0^-$, and it makes sense to apply (\ref{bracketdef}) for $\Phi = b_0^- \chi$, giving $\langle\chi |[\Psi^{\otimes n}]\rangle =\{b_0^-\chi\otimes \Psi^{\otimes n}\}$.} 

The BV master equation (\ref{bvmaster}) can be expressed more concisely using the string bracket as
\ie
n \{ Q_B \Psi\otimes \Psi^{\otimes (n-1)} \} = -  {1\over 2} \sum_{\ell=2}^{n-2} {n\choose \ell} \{ \Psi^{\otimes\ell} \otimes [\Psi^{\otimes (n-\ell)}] \} .
\fe
This is equivalent to the following property of $Q_B$ acting on the string bracket,\footnote{To derive this property requires a subtle formula $\langle \Phi|Q_B c_0^- | \Psi \rangle = -\langle\Phi |c_0^- Q_B|\Psi\rangle $ for $\Phi, \Psi\in\hat{\cal H}$. Note that $Q_B$ does not anti-commute with $c_0^-$. However, we have the manipulation
\ie
\langle \Phi|Q_B c_0^- | \Psi \rangle &= \langle \Phi|(b_0^-c_0^- + c_0^- b_0^-)Q_B c_0^- |\Psi\rangle
= \langle \Phi|c_0^- b_0^- Q_B c_0^- |\Psi\rangle
\\
&= \langle \Phi|c_0^- (L_0^- -  Q_B b_0^-) c_0^- |\Psi\rangle
= -\langle \Phi| c_0^- Q_B b_0^- c_0^- |\Psi\rangle
= -\langle \Phi| c_0^- Q_B |\Psi\rangle,
\fe
where we have repeatedly used the assumption that $\Phi$, $\Psi$ are annihilated by $b_0^-$ and $L_0^-$, yielding the claim.}
\ie\label{qbdistr}
-Q_B[\Psi^{\otimes n}] = n [Q_B \Psi\otimes \Psi^{\otimes (n-1)} ] + \sum_{\ell=1}^{n-2} {n\choose \ell} [\Psi^{\otimes \ell}\otimes [\Psi^{\otimes(n-\ell)}]].
\fe

\subsection{The string field equation}
\label{sec:sfteom}

The equation of motion for the string field $\Psi$ can be expressed concisely using the string field bracket as
\ie\label{sfteom}
Q_B \Psi + \sum_{n=2}^\infty {1\over n!} [\Psi^{\otimes n}] = 0.
\fe
It follows from the BV master equation (\ref{bvmaster}) or equivalently (\ref{qbdistr}) that (\ref{sfteom}) is invariant under the gauge transformation
\ie
\delta_\Lambda \Psi = Q_B \Lambda + \sum_{m=1}^\infty {1\over m!} [\Psi^{\otimes m}\otimes \Lambda],
\fe
where $\Lambda\in\hat{\cal H}$ plays the role of the infinitesimal gauge parameter.

An infinitesimal fluctuation of the string field solution from $\Psi$ to $\Psi+\delta \Psi$ is such that $\delta\Psi$ obeys the linearized SFT equation
\ie\label{linearizedsft}
Q_\Psi \delta\Psi \equiv Q_B \Psi + {1\over n!} \sum_{n=1}^\infty [\Psi^{\otimes n}\otimes \delta\Psi] = 0,
\fe
It follows from (\ref{qbdistr}) the ``deformed BRST charge'' $Q_\Psi$ obeys
\ie
Q_\Psi^2=0, 
\fe
and the gauge redundancy of $\delta\Psi$ can be simply expressed as $\delta\Psi \sim \delta \Psi + Q_\Psi \Lambda$. In other words, the gauge-inequivalent on-shell fluctuations of the string field correspond to the cohomology of $Q_\Psi$.

In practice, to find explicit solutions to the SFT equation typically requires fixing a gauge. A standard choice is the Siegel gauge $b_0^+\Psi = 0$. For our purpose, it can be convenient to work with a slightly relaxed gauge codnition
\ie\label{relaxedsiegel}
b_0^+ (1-\mathbb{P}) \Psi = 0.
\fe
It is useful to split the string field $\Psi$ according to the projector $\mathbb{P}^+$ as
\ie
\Psi = W + (1-\mathbb{P}^+) \Psi,~~~~ W \equiv \mathbb{P}^+ \Psi,
\fe
and rewrite (\ref{sfteom}) as
\ie\label{sfteomsplit}
& Q_B W + \mathbb{P}^+\sum_{n=2}^\infty {1\over n!}  [\Psi^{\otimes n}] = 0,
\\
& Q_B (1-\mathbb{P}^+) \Psi + (1-\mathbb{P}^+)\sum_{n=2}^\infty {1\over n!}  [\Psi^{\otimes n}] = 0.
\fe
Acting on the second equation with $b_0^+$, using $\{Q_B, b_0^+\} = L_0^+$ and that $L_0^+$ is invertible on the image of $1-\mathbb{P}^+$, we have
\ie\label{nonzerocomp}
(1-\mathbb{P}^+) \Psi = - {b_0^+\over L_0^+}(1-\mathbb{P}^+)  \sum_{n=2}^\infty {1\over n!} [\Psi^{\otimes n}]
\fe
In perturbation theory, one can solve $(1-\mathbb{P}^+)\Psi$ iteratively using (\ref{nonzerocomp}), provided that there is no cohomological obstruction to solving the first equation of (\ref{sfteomsplit}), i.e. provided that
\ie
\mathbb{P}^+\sum_{n=2}^\infty {1\over n!}  [\Psi^{\otimes n}]
\fe
is $Q_B$-exact, at each order in the expansion parameter.

\subsection{Asymmetric vertices and the flat-vertex string field frame}
\label{sec:flatvert}

It is possible to consider a deformed version of the $n$-string bracket, denoted by $[\cdot]_s$ where $s$ is a parameter, that is graded symmetric $n$-linear and still obey the analog of (\ref{qbdistr}), namely
\ie
-Q_B[\Psi^{\otimes n}]_s = n [Q_B \Psi\otimes \Psi^{\otimes (n-1)} ]_s + \sum_{\ell=1}^{n-2} {n\choose \ell} [\Psi^{\otimes \ell}\otimes [\Psi^{\otimes(n-\ell)}]_s]_s,
\fe
that does not come from a symmetric $(n+1)$-point string vertex, but rather an ``asymmetric vertex'' where one string field is regarded as outgoing and the remaining $n$ string fields incoming. Namely, for any $\Phi\in\hat{\cal H}$, we have
\ie
\langle \Phi| c_0^- |[\Psi^{\otimes n}]_s\rangle = {1\over (-2\pi i)^{n-2}} \int_{\Gamma_{n+1},s} \left\langle e^{\cal B} [ \Phi(0) ]^{f_1}  \prod_{i=2}^{n+1} [\Psi(0)]^{f_i}   \right\rangle
\fe
where $\Gamma_{n+1, s}$ is symmetric with respect to only the last $n$ punctures, and obeys an analog of the geometric master equation,
\ie
\partial \Gamma_{n+1,s} = -{1\over 2} \sum_{n_1+n_2=n+1} {n! \over n_1! n_2!} \sum_{i=2}^{n_1+1} \left[\hat\#_{i,1}(\Gamma_{n_1+1,s} \times \Gamma_{n_2+1,s}) \right]^{{\rm Sym}'},
\fe
where ${\rm Sym}'$ stands for symmetrization with respect to all but the first puncture.
Assuming $\Gamma_{n+1,s}$ depends smoothly on the deformation parameter $s$, the variation of the deformed string field bracket with respect to $s$ can be put in the form
\ie
\partial_s [\Psi^{\otimes n}]_s = Q_B \llbracket \Psi^{\otimes n} \rrbracket_s + \sum_{\ell=1}^{n-2} {n\choose \ell} \left( \llbracket \Psi^{\otimes \ell}\otimes [\Psi^{\otimes(n-\ell)}]_s \rrbracket_s -  [\Psi^{\otimes \ell}\otimes \llbracket\Psi^{\otimes(n-\ell)}\rrbracket_s]_s \right).
\fe
Here $\llbracket\Psi^{\otimes n}\rrbracket_s$ is the string field defined by
\ie
\langle \Phi| c_0^- | \llbracket\Psi^{\otimes n}\rrbracket_s\rangle = {1\over (-2\pi i)^{n-2}} \int_{\Gamma_{n+1},s} \left\langle {\cal B}_s e^{\cal B} [ \Phi(0) ]^{f_1}  \prod_{i=2}^{n+1} [\Psi(0)]^{f_i}   \right\rangle
\fe
for any $\Phi\in\hat{\cal H}$, where ${\cal B}_s$ is a contour integral of the $b$ ghost coming from the variation of the coordinate maps $f_i$ with respect to $s$,
\ie
{\cal B}_s = -\sum_{i=1}^n \oint_{\partial D_i} {dz\over 2\pi i} \partial_s f_i(w_i; t; s) b(z) +c.c.
\fe

Suppose $\Gamma_{n+1,s}$ is a family of asymmetric vertices, such that they agree with the symmetric vertex $\Gamma_{n+1}$ at $s=0$. Starting from a solution $\Psi$ of the SFT equation (\ref{sfteom}), we can then construct a family of string fields $\Psi(s)$ with \cite{Hata:1993gf}
\ie\label{dphis}
\partial_s \Psi(s) = \sum_{n=2}^\infty {1\over n!} \llbracket \Psi^{\otimes n}\rrbracket_s,
\fe
such that $\Psi(s)$ obeys an equation analogous to (\ref{sfteom}) but with the deformed string bracket,
\ie\label{deformsfteom}
Q_B \Psi(s) + \sum_{n=2}^\infty {1\over n!} [\Psi(s)^{\otimes n}]_s = 0.
\fe
While the asymmetric vertices themselves are not the Feynman vertices coming from an action functional, we see that the  equation (\ref{deformsfteom}) defined through the deformed bracket is in fact equivalent to the SFT equation  (\ref{sfteom}) up to the nonlinear field redefinition relating $\Psi(s)$ to $\Psi$. 

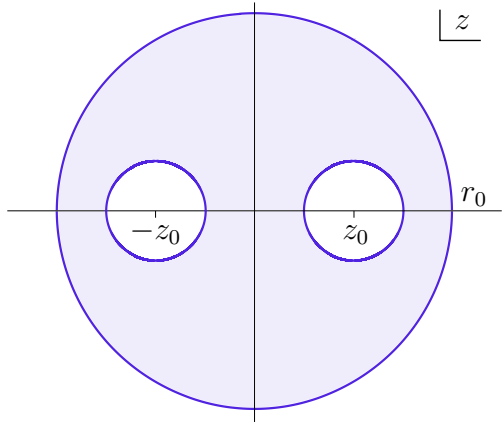
\begin{figure}[h!]
	\def\svgwidth{0.4\linewidth}
	\centering{
		%% Creator: Inkscape 1.1.2 (b8e25be8, 2022-02-05), www.inkscape.org
%% Accompanies image file 'FlatVertex.pdf' (pdf, eps, ps)
\begingroup%
  \makeatletter%
  \providecommand\color[2][]{%
    \errmessage{(Inkscape) Color is used for the text in Inkscape, but the package 'color.sty' is not loaded}%
    \renewcommand\color[2][]{}%
  }%
  \providecommand\transparent[1]{%
    \errmessage{(Inkscape) Transparency is used (non-zero) for the text in Inkscape, but the package 'transparent.sty' is not loaded}%
    \renewcommand\transparent[1]{}%
  }%
  \providecommand\rotatebox[2]{#2}%
  \newcommand*\fsize{\dimexpr\f@size pt\relax}%
  \newcommand*\lineheight[1]{\fontsize{\fsize}{#1\fsize}\selectfont}%
  \ifx\svgwidth\undefined%
    \setlength{\unitlength}{540bp}%
    \ifx\svgscale\undefined%
      \relax%
    \else%
      \setlength{\unitlength}{\unitlength * \real{\svgscale}}%
    \fi%
  \else%
    \setlength{\unitlength}{\svgwidth}%
  \fi%
  \global\let\svgwidth\undefined%
  \global\let\svgscale\undefined%
  \makeatother%
  \begin{picture}(1,0.87037037)%
    \lineheight{1}%
    \setlength\tabcolsep{0pt}%
    \put(0,0){\includegraphics[width=\unitlength,page=1]{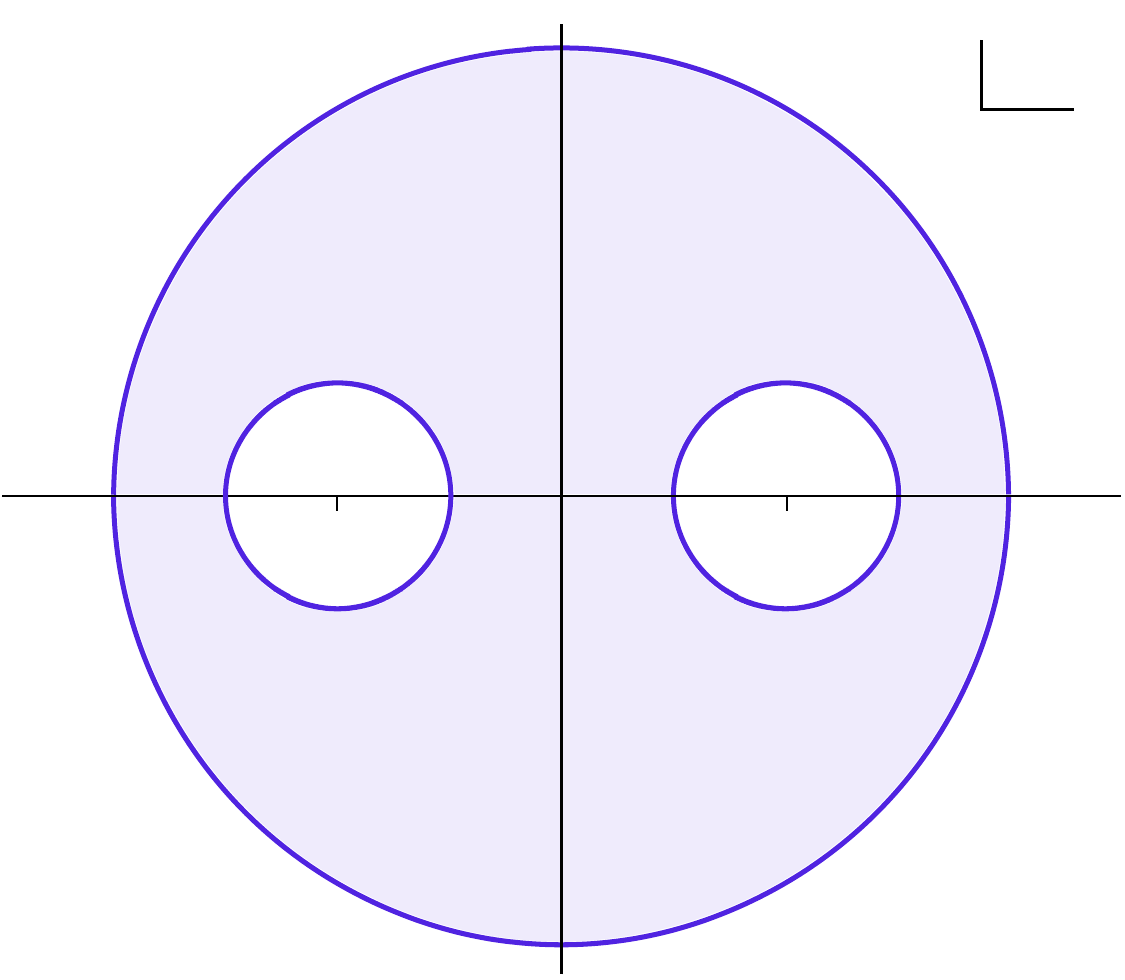}}%
    \put(0.90168321,0.79673083){\makebox(0,0)[lt]{\lineheight{1.25}\smash{\begin{tabular}[t]{l}$z$\end{tabular}}}}%
    \put(0.67611598,0.37235379){\makebox(0,0)[lt]{\lineheight{1.25}\smash{\begin{tabular}[t]{l}$z_0$\end{tabular}}}}%
    \put(0.24797849,0.37235379){\makebox(0,0)[lt]{\lineheight{1.25}\smash{\begin{tabular}[t]{l}$-z_0$\end{tabular}}}}%
    \put(0.9146273,0.44890329){\makebox(0,0)[lt]{\lineheight{1.25}\smash{\begin{tabular}[t]{l}$r_0$\end{tabular}}}}%
  \end{picture}%
\endgroup%
	\caption{ The asymmetric flat 3-vertex, with three local charts covering the unshaded regions. The outgoing string field is inserted at infinity, whereas the incoming string fields are inserted at $-z_0$ and $z_0$ respectively.
			\label{fig:flatvertex}
	}}
\end{figure}

In practice, asymmetric vertices can be constructed more simply than symmetric ones, and can be employed to simplify the calculations in solving the SFT equation and analyzing the spectrum of fluctuations. One particularly convenient choice, of which we will make use extensively in this paper, is the ``flat'' asymmetric vertices whose coordinate maps $f_2,\cdots, f_n$ (for the incoming string fields) are composed only of rotations, scalings, and translations on the complex plane, whereas $f_1$ (for the outgoing string field) involves only inversion. For instance, we can deform the symmetric ${\rm PSL}(2,\mathbb{C})$ vertex $\Gamma_3$ to an asymmetric flat vertex $\Gamma_{3,s=1}$, whose coordinate maps are
\ie
f_1(w_1) = r_0 w_1^{-1}, ~~~ f_2(w_2) = w_2-z_0,~~~ f_3(w_3) = w_3 +z_0,
\fe
where $r_0\geq |z_0|+1$, $|z_0|\geq 1$, as portrayed in Figure \ref{fig:flatvertex}. The resulting 2-string ``flat'' bracket is simply
\ie\label{twostringflat}
[\Psi^{\otimes 2}]_1 = b_0^-\mathbb{P}^- r_0^{-L_0^+} \left(\Psi(-z_0) \Psi(z_0) \right).
\fe
Compatible higher point vertices can be constructed similarly (an explicit asymmetric flat 4-vertex will be used in section \ref{sec:ccthird}). We will refer to the corresponding deformed string field $\Psi(1)$ as the {\it flat-vertex frame} string field.

\section{SFT solutions for CFT deformations}
\label{sec:pertsol}

\subsection{The general logic of background independence}

Given two different worldsheet CFTs ${\mathfrak A}$ and ${\mathfrak B}$, and a choice of string vertices as chains in $\hat{\cal P}_{0,n}$ that obey the geometric master equation, one constructs a priori two different classical closed SFTs, which we refer to as SFT$_{\mathfrak A}$ and SFT$_{\mathfrak B}$. The corresponding space of string fields will be denoted $\hat {\cal H}_{\mathfrak A}$ and $\hat{\cal H}_{\mathfrak B}$ respectively.

At least when $\mathfrak A$ and $\mathfrak B$ are connected by a continuous exactly marginal deformation, and possibly more generally, we expect the change of background ${\mathfrak A}\to{\mathfrak B}$ to be equivalently represented by a nontrivial string field in SFT$_{\mathfrak A}$,
\ie
\Psi_{{\mathfrak A} | {\mathfrak B}} \in \hat {\cal H}_{\mathfrak A},
\fe
that satisfies its equation of motion. The idea of ``background independence'' of SFT is that SFT$_{\mathfrak A}$ and SFT$_{\mathfrak B}$ should be equivalent up to a field redefinition
\ie
F_{\mathfrak{A}, \mathfrak{B}}: {\cal H}_{\mathfrak A} \to {\cal H}_{\mathfrak B},
\fe
that in particular maps $\Psi_{{\mathfrak A} | {\mathfrak B}}$ to the vacuum field configuration in SFT$_{\mathfrak B}$, namely
\ie
F_{\mathfrak{A}, \mathfrak{B}}(\Psi_{{\mathfrak A} | {\mathfrak B}}) = 0.
\fe
Furthermore, on-shell observables in the background ${\mathfrak B}$ such as the string mass spectrum or more generally dispersion relations, as well as the string scattering amplitudes, should be reproduced in SFT$_{\mathfrak A}$ through fluctuations around the background solution $\Psi_{{\mathfrak A}|{\mathfrak B}}$. This is proven for infinitesimal marginal deformations in \cite{Sen:1990hh, Sen:1990na, Sen:1992pw, Sen:1993mh, Sen:1993kb}.

By the same token, given a solution $\Psi\in \hat{\cal H}_{\mathfrak A}$, one may postulate that $\Psi = \Psi_{{\mathfrak A}|{\mathfrak B}}$ for some a priori unspecified worldsheet CFT $\mathfrak B$, and try to recover the data of CFT ${\mathfrak B}$ from on-shell observables associated with fluctuations around $\Psi$ in SFT$_{\mathfrak A}$. We will employ this idea both to construct SFT solutions (e.g. that describe closed string tachyon condensation), and to use SFT to formulate a manifestly finite form of (RG-improved) conformal perturbation theory of general CFTs.\footnote{The analog of this approach for boundary CFT and open SFT was previously explored in \cite{Budzik:2020aqg, Scheinpflug:2023osi}.}

\subsection{Embedding a general CFT into SFT}

Let $\mathbb{A}$ be a two-dimensional CFT of central charge $c_{\mathbb A}$. In the most basic setup, we embed $\mathbb{A}$ into a worldsheet CFT of the form
\ie\label{amcft}
\mathbb{A}\otimes \mathbb{M} \otimes (bc {\rm ~system})
\fe
where $\mathbb{M}$ is an auxiliary CFT of central charge $26-c_{\mathbb{A}}$. For the consideration of this paper, $\mathbb{M}$ need not be unitary nor modular invariant, but will be assumed to contain the identity operator and formal Virasoro ``mock'' primaries $\Upsilon_{h,\widetilde h}$ of generic weights $(h, \widetilde h)$, subject to only the condition $h-\widetilde h \in \mathbb{Z}$.

Heuristically, $\mathbb{A}$ may be viewed as the internal CFT that describes a string ``compactification'', even though $\mathbb{A}$ need not be a compact CFT. The auxiliary CFT $\mathbb{M}$, on the other hand, imitates the noncompact ``Minkowskian-like'' spacetime in which the string modes propagate and scatter, even though $26-c_{\mathbb A}$ need not be a positive integer and we need not commit to a given set of structure constants for the mock primaries.

A deformation of the CFT $\mathbb{A}$ to $\mathbb{A}_\lambda$ (where $\lambda$ stands for the deformation parameter) that preserves its central charge would lead to a deformed worldsheet CFT
\ie\label{newamcft}
\mathbb{A}_\lambda\otimes \mathbb{M} \otimes (bc {\rm ~system}),
\fe
representing a new spacetime background where the internal space is deformed. From the perspective of the classical bosonic closed SFT based on the original worldsheet CFT (\ref{amcft}), the new background may be equivalently represented by a string field
\ie\label{psilambhc}
\Psi_\lambda \in \hat{\cal H}_0 \subset {\cal H}_\mathbb{A}\otimes {\cal V}_0 \otimes {\cal H}_{bc}
\fe
that solves the SFT equation (\ref{sfteom}). Here ${\cal V}_0$ is the space of $c=26-c_{\mathbb A}$ Virasoro descendants of the identity operator in the auxiliary CFT $\mathbb{M}$, and $\hat{\cal H}_0$ is the subspace of ${\cal H}_\mathbb{A}\otimes {\cal V}_0 \otimes {\cal H}_{bc}$ annihilated by $b_0^-, L_0^-$. Note that the non-identity primaries of $\mathbb{M}$ do not play any role in constructing the solution $\Psi_\lambda$ itself, and that the SFT equation only involves CFT data of the original undeformed CFT $\mathbb{A}$. To extract CFT data of $\mathbb{A}_\lambda$, as we will illustrate in section \ref{sec:recover}, requires studying fluctuations around $\Psi_\lambda$, where generic mock primaries of $\mathbb{M}$ would be turned on.

\subsection{Perturbative SFT solution for marginal deformation}
\label{sec:marginal}

We begin by considering the deformation of CFT $\mathbb{A}$ by a marginal operator, i.e. a weight $(1,1)$ Virasoro primary ${\cal O}$, by inserting $\exp(-\Delta S)$ into correlation functions, where
\ie
\Delta S = -{\lambda\over\pi} \int d^2z {\cal O}(z,\bar z).
\fe
Such a deformation is typically defined with a regularization scheme that cuts off the UV divergence when the deformation operators collide with one another, e.g. by removing the region of integration where the distance between a pair of deformation operators is less than $\epsilon$, and a renormalization of the coupling constant $\lambda$ may be necessary in taking the $\epsilon\to 0$ limit. When the deformation is exactly marginal, all UV divergences involve inverse powers of $\epsilon$ and can be subtracted off in a way that is consistent with locality, up to the ambiguity of a finite renormalization of $\lambda$, whereas logarithmic divergences that would lead to a nontrivial RG flow are absent. In this case, the deformation results in a 1-parameter family of CFTs $\mathbb{A}_\lambda$.

Via embedding into SFT as described in the previous subsection, we can associate an exactly marginal deformation with a perturbative solution to the SFT equation, of the form
\ie\label{marginalexpan}
\Psi_\lambda = \sum_{n=1}^\infty \lambda^n \Psi_n,
\fe
where $\Psi_n\in \hat{\cal H}_0$ (defined as in (\ref{psilambhc})), and the first order string field involves the deformation operator ${\cal O}$, 
\ie\label{psiona}
\Psi_1 = c\widetilde c {\cal O}.
\fe
Indeed, when ${\cal O}$ is a marginal primary, $\Psi_1$ is $Q_B$-closed, and the SFT equation is satisfied to first order in $\lambda$. One can proceed to find the higher order terms $\Psi_n$ in the gauge (\ref{relaxedsiegel}) by the procedure outlined in section \ref{sec:sfteom}, with the recursive formula
\ie\label{psinwn}
\Psi_n = W_n - {b_0^+\over L_0^+} (1-\mathbb{P}^+) \sum_{k=2}^n\sum_{\substack{ \{n_1, \cdots, n_k\} \\ n_1+\cdots+n_k = n} } {1\over \prod_{i=1}^k n_i!} \left[ \bigotimes_{i=1}^k \Psi_{n_i} \right],
\fe
where $W_n$ obeys
\ie\label{qbwrec}
Q_B W_n = - \mathbb{P}^+\sum_{k=2}^n\sum_{\substack{ \{n_1, \cdots, n_k\} \\ n_1+\cdots+n_k = n} } {1\over \prod_{i=1}^k n_i!} \left[ \bigotimes_{i=1}^k \Psi_{n_i} \right] \equiv \chi_n.
\fe
If we have solved $\Psi_1, \cdots, \Psi_{n-1}$, the RHS of (\ref{qbwrec}) which we denote by $\chi_n$ is necessarily $Q_B$-closed by virtue of (\ref{qbdistr}). However, if $\chi_n$ represents a nontrivial $Q_B$-cohomology, it would present an obstruction to solving the SFT equation at order $n$.

When the CFT $\mathbb A$ is compact and unitary, a complete basis of the $Q_B$-cohomology in $\hat{\cal H}_0\subset {\cal H}_{\mathbb A}\otimes {\cal V}_0 \otimes {\cal H}_{bc}$ is represented by the following states:

~~~~~~\begin{tabular}{c||c|c|c|c|c|c}
	$gh \#$ &0 & 1 & 2 & 3 & 4 & 5 \\ 
	\hline 
	state & $1$ &
	\begin{tabular}{@{}c@{}} $cj_a$ \\ $\tilde c \tilde j_a$ \end{tabular}& 
	\begin{tabular}{@{}c@{}} $c \partial^2 c - \tilde c \bar \partial^2 \tilde c$ \\ $c \tilde c \mc O_i$\end{tabular} &
	\begin{tabular}{@{}c@{}} $c_0^+ \left(c\partial^2 c - \tilde c \bar \partial^2 \tilde c\right)$\\ $c_0^+ c \tilde c \mc O_i$ \end{tabular}&
	\begin{tabular}{@{}c@{}}$c_0^+ \tilde c \bar \partial^2 \tilde c c j_a$ \\ $c_0^+ c \partial^2 c \tilde c \tilde j_a$ \end{tabular}&
	$c_0^+ c \partial^2 c \tilde c \bar \partial^2 \tilde c$
\end{tabular}

\noindent where $j_a, \widetilde j_a$ are holomorphic and anti-holomorphic (weight 1) currents, and ${\cal O}_i$ are weight $(1,1)$ primaries. The physical string field is restricted to ghost number 2, whereas potential cohomological obstructions to solving (\ref{qbwrec}) arises at ghost number 3. There are two types of obstructions that may appear in $\chi_n$: the appearance of $c_0^+ c \tilde c \mc O_i$ would represent an order $\lambda^n$ contribution to the RG beta function that violates marginality, whereas the appearance of the ``anti-ghost dilaton'' operator $c_0^+(c \partial^2 c - \tilde c \bar \partial^2 \tilde c)$ would represent an order $\lambda^n$ shift of the central charge that violates criticality of the worldsheet theory.

\subsection{The role of dilaton and central charge shift}
\label{sec:ccshift}

For a noncompact CFT $\mathbb{A}$, an exactly marginal deformation can deform the central charge. When embedded in the worldsheet CFT, a central charge deformation would violate criticality. In the SFT language, this corresponds to an obstruction to solving the SFT equation by a term proportional to the anti-ghost dilaton $c_0^+(c \partial^2 c - \tilde c \bar \partial^2 \tilde c)$.

It is useful to illustrate this phenomenon in the example of the linear dilaton deformation. Let $\mathbb{D}_\beta$ be the linear dilaton CFT described by the scalar field $Y$ and the stress-energy tensor
\ie
T = - {1\over 2} (\partial Y)^2 + {\B\over 2} \partial^2 Y,
\fe
where $\beta$ is the background charge, and the central charge is $c_{{\mathbb D}_\beta} = 1+ 3\B^2$.\footnote{Our convention is such that $Y(z, \bar z) Y(0) \sim - \log|z|^2$, with $L_1\partial Y = \beta$, and $L_0 Y = {\B\over 2}$.} The worldsheet matter CFT is taken to the tensor product of $\mathbb{D}_\beta$ with an auxiliary CFT of central charge $26-c_{\mathbb{D}_\beta}$.

An infinitesimal deformation of the background charge is represented by a string field 
\ie
\Psi = \epsilon \Theta + {\cal O}(\epsilon^2),
\fe
where %\footnote{An alternative, BRST-equivalent, expression for $\Theta$ is $c \partial^2 c Y - 2 c_0^+ \tilde c \bar \partial Y$ or $- \tilde c \bar \partial^2 \tilde c Y + 2 c_0^+ c \partial Y$.}
\ie\label{dexpr}
\Theta = {1\over 2} (c\partial^2c - \wt c \bar\partial^2\wt c) Y + c_0^+ (c \partial Y - \wt c \bar\partial Y)
\fe
satisfies
\ie\label{qbdexpr}
Q_B \Theta = \B c_0^+ (c\partial^2 c - \wt c \bar\partial^2\wt c).
\fe
That is, $\Psi = \epsilon \Theta$ violates the first order string field equation by the aforementioned anti-ghost dilaton obstruction term, which reflects a shift of central charge of the linear dilaton CFT due to the background charge deformation. 

Note that the expression (\ref{dexpr}) for $\Theta$ is not in the Siegel gauge. A BRST-equivalent Siegel gauge representative is $\Theta' = b_0^+ Q_B \chi$, where $\chi$ obeys $L_0^+\chi = \Theta$. One can verify that in the $\beta\to 0$ limit, $\Theta'$ reduces to the more familiar expression for the vertex operator of a linear dilaton profile \cite{Bergman:1994qq}, $\Theta'|_{\beta=0} = -2 c\wt c Y \partial Y \bar\partial Y +  (c\partial^2c - \wt c \bar\partial^2\wt c) Y$. Indeed, such a deformation shifts the background charge of the linear dilaton CFT by $\Delta\beta = -2\epsilon$, which leads to a first order shift of central charge by 
\ie
\Delta c =12\B\epsilon.
\fe

While the deformation of the linear dilaton background charge by itself does not correspond to a SFT solution, we observe in (\ref{qbdexpr}) that the BRST cohomology of the anti-ghost dilaton is trivialized in the presence of the linear dilaton CFT. This motivates embedding the CFT $\mathbb{A}$ into a worldsheet CFT of the form
\ie
\mathbb{A}\otimes \mathbb{D}_\beta \otimes \mathbb{M}\otimes (bc{\rm ~system}),
\fe
where the auxiliary CFT has central charge $26-c_{\mathbb A}-c_{\mathbb{D}_\beta}$, and consider a string field solution
\ie\label{psiadm}
\Psi_\lambda\in \hat{\cal H}_0 \subset {\cal H}_{\mathbb A}\otimes {\cal H}_{\mathbb{D}_\beta} \otimes {\cal V}_0\otimes {\cal H}_{bc}
\fe
that corresponds to the deformed worldsheet CFT
\ie\label{admfact}
\mathbb{A}_\lambda\otimes \mathbb{D}_{\beta(\lambda)} \otimes \mathbb{M}\otimes (bc{\rm ~system}),
\fe
where $c_{{\mathbb A}_\lambda}$ may differ from $c_{\mathbb A}$, but this is compensated by a deformation of the background charge of the linear dilaton sector to $\beta(\lambda)$, so that $c_{{\mathbb A}_\lambda} + c_{\mathbb{D}_{\beta(\lambda)}} = c_{\mathbb A} +  c_{\mathbb{D}_\beta}$. In SFT description, starting with a first order string field of the form (\ref{psiona}), a shift of the central charge of $\mathbb{A}$ which corresponds to an anti-ghost dilaton contribution on the RHS of (\ref{qbwrec}) would require turning on a dilaton string field proportional to $\Theta$ (\ref{qbdexpr}) in $W_n$.

A general string field solution of the form (\ref{psiadm}) may couple the linear dilaton sector nontrivially to $\mathbb{A}$. For the purpose of analyzing $\mathbb{A}_\lambda$, we would like to ensure that the SFT solution corresponds to the factorized worldsheet theory of the form (\ref{admfact}). In principle, this can be checked from the spectrum and amplitudes of the fluctuations around $\Psi_\lambda$. A particularly simple diagnosis of a decoupled linear dilaton sector $\mathbb{D}_{\beta'}$ is the existence of the shift current $j = \partial Y'$, which is an ``almost Virasoro-primary'' in that $j$ is annihilated by $L_{n\geq 2}$ but with $L_1 j = \beta'$ (and $L_0 j = 0$). In the SFT description, this corresponds to linearized fluctuation string field ${\cal J}_\Psi$ of ghost number 1  around the background solution $\Psi$ that obeys
\ie\label{jpsicri}
& Q_\Psi {\cal J}_\Psi = -\beta' {\cal K}_\Psi.
\fe
where $Q_\Psi$ is the deformed BRST charge as defined in (\ref{linearizedsft}), and ${\cal K}_\Psi$ is the ghost number 2 string field that corresponds to a constant shift of the dilaton. In the undeformed background, ${\cal J}_\Psi$ and ${\cal K}_{\Psi}$ reduce to
\ie\label{jkzero}
{\cal J}_0 = c\partial Y - \wt c \bar\partial Y,~~~~ {\cal K}_0 = {1\over 2} (c\partial^2 c - \wt c \bar\partial^2 \wt c),
\fe
where ${\cal K}_0$ is also known as the ghost dilaton operator \cite{Bergman:1994qq}. Further aspects of this analysis will be discussed in section \ref{sec:decoupling}.

\section{SFT solution for short RG flows}

The perturbative SFT solutions that describe marginal deformations, as outlined in the previous section, can be generalized to describe a slightly relevant deformation of the CFT $\mathbb{A}$ that generates a short RG flow to a new fixed point CFT $\mathbb{A}_\lambda$ \cite{Mukherji:1991tb}.  Here, a renormalized coupling $\lambda$ can be identified with the expansion parameter used in constructing the string field $\Psi_\lambda$, and the IR fixed point value of $\lambda$ will determined by the SFT equation.

Let us assume that the CFT $\mathbb{A}$ of interest admits a slightly relevant operator $\phi$ of weight $h=\wt h = 1-\varepsilon$, where $\varepsilon$ is a small parameter that also controls the length of the RG flow generated by $\phi$, in the sense that a suitable renormalized coupling will be small at the RG fixed point. We will embed $\mathbb{A}$ into a worldsheet theory of the form (\ref{admfact}), and denote by $\mathbb{P}^+_\varepsilon$ the projector from the space of string fields $\hat{\cal H}$ to a subspace on which $|L_0^+| \lesssim {\cal O}(\varepsilon)$.

\subsection{A perturbative SFT solution describing the RG fixed point}
\label{sec:pertrg}

Working in the gauge
\ie
b_0^+(1-\mathbb{P}^+_\varepsilon) \Psi = 0,
\fe
and writing 
\ie
\Psi = W + \Psi_>,~~~~ W\equiv \mathbb{P}_\varepsilon^+\Psi,~~~\Psi_>\equiv (1-\mathbb{P}_\varepsilon^+) \Psi,
\fe
we can construct a solution by iteratively solve
\ie{}
& Q_B W = - \mathbb{P}_\varepsilon^+ \sum_{n\geq 2} {1\over n!} \left[ (W + \Psi_>)^{\otimes n}\right],
\\
& \Psi_> = - {b_0^+\over L_0^+} (1-\mathbb{P}_\varepsilon^+) \sum_{n\geq 2} {1\over n!} \left[ (W + \Psi_>)^{\otimes n}\right].
\fe
We can further eliminate $\Psi_>$ and write an equation for $W$ itself, of the form
\ie\label{weqn}
Q_B W = - {1\over 2} \mathbb{P}_\varepsilon^+ [W^{\otimes 2}] + {1\over 2}  \mathbb{P}_\varepsilon^+ \Big[W \otimes \Big( {b_0^+\over L_0^+} (1-\mathbb{P}_\varepsilon^+) [W^{\otimes 2}] \Big)\Big] - {1\over 6} \mathbb{P}_\varepsilon^+ [W^{\otimes 3}]  + \cdots
\fe
To proceed, we consider a perturbative expansion
\ie
W = \sum_{n=1}^\infty \varepsilon^n W_n,
\fe
where $W_1$ takes the form of a first order deformation by the relevant operator $\phi$, 
\ie
W_1 = g_1 c \wt c \phi.
\fe
Note that $W_1$ is not BRST closed, since $\phi$ is not marginal. However, $Q_B W_1$ is of order $\varepsilon$ due to $\phi$ being nearly marginal,
\ie
Q_B W_1 = - 2\varepsilon g_1 c_0^+ c\wt c\phi.
\fe
The SFT equation (\ref{weqn}) at order $\varepsilon^2$ gives
\ie\label{sftsecondor}
-2 \varepsilon^2 g_1 c_0^+ c\wt c \phi + \varepsilon^2 Q_B W_2 = - {\varepsilon^2 g_1^2\over 2} \mathbb{P}_\varepsilon^+ \left[c\wt c \phi\otimes c\wt c \phi \right] + {\cal O}(\varepsilon^3).
\fe
To carry out the explicit calculations, it will be convenient to work in the flat-vertex frame defined in section \ref{sec:flatvert},\footnote{The reader should be cautious that the string field solution in the flat-vertex frame, which solves the deformed SFT equation of the form (\ref{deformsfteom}), is related to the string field that appears in an action (necessarily constructed from symmetric vertices) by the field redefinition (\ref{dphis}).} in which the flat 2-string bracket (\ref{twostringflat}) gives
\ie\label{flatctwo}
\mathbb{P}_\varepsilon^+ \left[c\wt c \phi\otimes c\wt c \phi \right] &= b_0^- \mathbb{P}^- r_0^{-L_0^+} \mathbb{P}_\varepsilon^+ (c\wt c \phi(-z_0) c\wt c \phi(z_0))
\\
&= 2 \left({r_0\over 2|z_0|}\right)^{2\varepsilon} C_{\phi\phi\phi} \, c_0^+ c \wt c  \phi,
\fe
where $C_{\phi\phi\phi}$ is the structure constant appearing in the $\phi \phi$ OPE, assuming that $\phi$ is the only nearly marginal operator. With $W_2 = g_2 c\wt c \phi$ for some order one coefficient $g_2$ and thus $Q_B W_2 = {\cal O}(\varepsilon)$, we can solve (\ref{sftsecondor}) by setting
\ie\label{gonefi}
g_1 = {2\over C_{\phi\phi\phi}}.
\fe
Therefore, we have determine $W$ up to second order to be of the form
\ie
W = \lambda c\wt c \phi + {\cal O}(\varepsilon^3),~~~~\lambda = {2\varepsilon\over C_{\phi\phi\phi}} + {\cal O}(\varepsilon^2).
\fe
$\lambda$ can be viewed as a renormalized coupling at the RG fixed point.
Note that in particular we have not invoked a term proportional to $\Theta$ (\ref{dexpr}) in the linear dilaton sector, indicating that there is no shift of central charge of the deformed CFT $\mathbb{A}_\lambda$ at order $\varepsilon^2$.

\subsection{Central charge shift at the third order}
\label{sec:ccthird}

\begin{figure}
	\def\svgwidth{0.7\linewidth}
	\centering{
		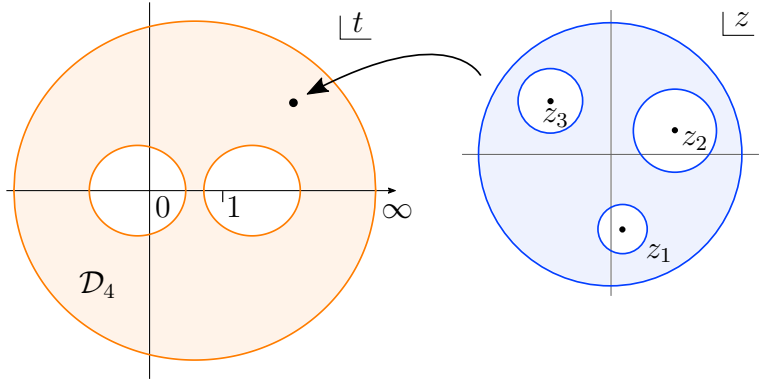	\caption{The domain ${\cal D}_4$ in the moduli space ${\cal M}_{0,4}$ parameterized by the cross ratio $t$, and corresponding 4-punctured sphere with the local charts of a flat 4-vertex.
			\label{d4figure}
	}}
\end{figure}

We seek an anti-ghost dilaton term of the form
\ie\label{antighosd}
a\, c_0^+(c\partial^2 c - \wt c \bar\partial^2\wt c)
\fe
on the RHS of (\ref{weqn}), which would require turning on a dilaton component of $W$ in the form $\B^{-1} a \Theta$, where $\Theta$ is given by (\ref{dexpr}), resulting in a shift of the central charge of $\mathbb{A}_\lambda$ by
\ie\label{carel}
\Delta c \equiv c_{\mathbb{A}_\lambda} - c_{\mathbb A} = -12 a.
\fe 
At order $\varepsilon^3$, the relevant terms on the RHS of  (\ref{weqn}) are\footnote{There are also order $\varepsilon^3$ terms coming from $-{1\over 2} \mathbb{P}_\varepsilon[W_1^{\otimes 2}]$, due to the $\varepsilon$-dependence of (\ref{flatctwo}), but these do not contain the anti-ghost dilaton. }
\ie\label{wonecub}
- \varepsilon^3 \mathbb{P}_\varepsilon^+ [W_1\otimes W_2] + {\varepsilon^3\over 2}  \mathbb{P}_\varepsilon^+ \Big[W_1 \otimes \Big( {b_0^+\over L_0^+} (1-\mathbb{P}_\varepsilon^+) [W_1^{\otimes 2}] \Big)\Big] - {\varepsilon^3\over 6} \mathbb{P}_\varepsilon^+ [W_1^{\otimes 3}].
\fe
Let $\sigma$ be the difference between the holomorphic and anti-holomorphic ghost numbers, e.g. $c$ carries $\sigma=1$ and $\wt c$ carries $\sigma=-1$. $W_1$ and $W_2$ both have $\sigma=0$. The 2-string bracket changes $\sigma$ by $\pm 1$, however $b_0^+[W_1^{\otimes 2}]$ has $\sigma=0$. Therefore, only the last term in (\ref{wonecub}) involving the 3-string bracket can contain an anti-ghost dilaton, which involves terms of $\sigma=\pm 3$. We adopt an asymmetric flat 4-string vertex with coordinate maps
\ie\label{coordmapexp}
&f_1(w_1) = r_0 w_1^{-1},
\\
& f_2(w_2) = q_1 w_2 + z_1,
\\
& f_3(w_3) = q_2 w_3 + z_2,
\\
& f_4(w_4) = q_3 w_4 + z_3,
\fe
where $q_i=q_i(t, \bar t)$ and $z_i=z_i(t, \bar t)$ are suitably chosen functions of the complex modulus $t$ over a domain ${\cal D}_4\subset {\cal M}_{0,4}$ (Figure \ref{d4figure}), such that on the three boundaries of the domain they match with the plumbing of a pair of flat 3-vertices modulo constant phase rotations of $w_i$ according to the geometric master equation. If we take $t$ to be the cross ratio $z_{31}/z_{21}$, the 4-vertex domain is
\ie
{\cal D}_4 = \left\{ t\in\mathbb{C}:~ \left|t-{1\over 2}\right|, \left|{1\over t}-{1\over 2}\right|, \left|{1\over 1-t} - {1\over 2} \right| <r_0 \right\}.
\fe

The matching condition on the boundary $t={1\over 2}+r_0 e^{i\theta}$ for instance is
\ie\label{matching}
& |q_1|=|q_2|={1\over r_0},~~~ |q_3| = 1,
\\
& z_1 = -z_0 {t+{1\over 2}\over t-{1\over 2}},~~~ z_2 = -z_0 {t-{3\over 2}\over t-{1\over 2}}, ~~~ z_3=z_0.
\fe
The resulting flat 3-string bracket is
\ie
[\Psi^{\otimes 3}] = {b_0^- \mathbb{P}^- \over (-2\pi i)}\int_{{\cal D}_4} dt\wedge d\bar t B_t B_{\bar t}\, (|q_1|^{L_0^+} \Psi)(z_1) (|q_2|^{L_0^+} \Psi)(z_2) (|q_3|^{L_0^+} \Psi)(z_3),
\fe
where
\ie\label{btexpr}
B_t = \sum_{i=1}^3 \left[ b_{-1}^{(z_i)} {\partial z_i\over \partial t} + b_0^{(z_i)} q_i^{-1}{\partial q_i\over \partial t} + \wt b_{-1}^{(\bar z_i)} {\partial \bar z_i\over \partial t} + \wt b_0^{(\bar z_i)} \bar q_i^{-1}{\partial \bar q_i\over \partial t}  \right].
\fe
Here $b_n^{(z_i)}$ stands for $b_n$ acting on the operator inserted at $z_i$.

The anti-ghost dilaton coefficient at order $\varepsilon^3$ can now be evaluated as
\ie\label{aintexp}
a &= {\varepsilon^3\over 6} b_0^+\mathbb{P}_\varepsilon^+ [W_1^{\otimes 3}]|_{\wt c\bar \partial^2 \wt c}
\\
&= - {\varepsilon^3 \over 6(2\pi i)} b_0^+ b_0^- \mathbb{P}_\varepsilon^+ \mathbb{P}^- \int_{{\cal D}_4} dt\wedge d\bar t \left(\sum_{i=1}^3 b_{-1}^{(z_i)} {\partial z_i\over \partial t}\right) \left(\sum_{j=1}^3 b_{-1}^{(z_j)} {\partial z_j\over \partial \bar t}\right) W_1(z_1) W_1(z_2) W_1(z_3) \bigg|_{\wt c\bar \partial^2 \wt c}
\\
&= - {\varepsilon^3g_1^3 C_{\phi\phi\phi} \over 6(2 \pi i)}  \int_{{\cal D}_4} \omega,
\fe
where
\ie
\omega &=   {z_3 dz_1\wedge dz_2 \over z_{12}z_{13}z_{23}} + {\rm cyclic} = d\zeta,~~~~\zeta={z_1+z_2+z_3\over 3} \left( {dz_1\over z_{12}z_{13}} + {\rm cyclic} \right).
\fe
(\ref{aintexp}) is thus reduced to a boundary integral, which can be evaluated using the matching condition (\ref{matching}) to give
\ie
a &= - {\varepsilon^3g_1^3 C_{\phi\phi\phi} \over 6(2 \pi i)}  \int_{\partial {\cal D}_4} \zeta
\\
&=  - {\varepsilon^3g_1^3 C_{\phi\phi\phi} \over 6(2 \pi i)} \cdot 3\int_{C_\infty} {(1-2t) dt\over 12 t(t-1)}
\\
&= {1\over 12} \varepsilon^3g_1^3 C_{\phi\phi\phi} .
\fe
Substituting the solution for $g_1$ (\ref{gonefi}) and using the relation (\ref{carel}), we find the central charge shift
\ie
\Delta c = -{8\varepsilon^3\over C_{\phi\phi\phi}^2}
\fe
in agreement with \cite{Zamolodchikov:1987ti, Mukherji:1991tb}.

\section{Recovering CFT data from SFT}
\label{sec:recover}

\subsection{CFT spectrum from string field fluctuations}
\label{sec:spec}

To recover the CFT data of $\mathbb{A}_\lambda$ that corresponds to a given SFT solution $\Psi_\lambda$ (\ref{psilambhc}), we begin by considering string field fluctuations around $\Psi_\lambda$ of the form
\ie\label{phihh}
\Phi_{h,\widetilde h} \in \hat{\cal H}_{h, \widetilde h}\subset {\cal H}_{\mathbb{A}} \otimes {\cal V}_{1-h, 1-\widetilde h}\otimes {\cal H}_{bc},
\fe
where ${\cal V}_{1-h, 1-\widetilde h}$ is the space of Virasoro descendants of the mock primary $\Upsilon_{1-h,1-\widetilde h}$ in the auxiliary CFTs, and $\hat{\cal H}_{h, \wt h}$ is the subspace of ${\cal H}_{\mathbb{A}} \otimes {\cal V}_{1-h, 1-\widetilde h} \otimes {\cal H}_{bc}$ annihilated by $b_0^-$ and $L_0^-$. We seek solutions to the linearized SFT equation
\ie\label{linearhh}
Q_{\Psi_\lambda} \Phi_{h, \widetilde h} = 0
\fe
of ghost number 2, where $Q_{\Psi_\lambda}$ is the deformed BRST charge defined as in (\ref{linearizedsft}), that represent on-shell fluctuations, subject to the gauge redundancy $\Phi_{h, \wt h} \sim \Phi_{h, \wt h}+Q_{\Psi_\lambda}\chi$. Note that by construction $\Psi_\lambda$ does not involve non-identity primaries of the auxiliary CFT, and therefore the equation (\ref{linearhh}) does not involve any CFT data of the auxiliary CFT other than the assumed conformal weight of the mock primary.

The background independence of SFT (proven for infinitesimal deformations in \cite{Sen:1990hh, Sen:1990na, Sen:1992pw, Sen:1993mh, Sen:1993kb}) would imply an isomorphism between the cohomology of $Q_{\Psi_\lambda}$ on $\hat{\cal H}_{h, \widetilde h}$ and the (undeformed) BRST cohomology of the deformed worldsheet CFT (\ref{newamcft}) restricted to the subspace of
\ie\label{alambdhh}
{\cal H}_{\mathbb{A}_\lambda}\otimes {\cal V}_{1-h, 1-\widetilde h} \otimes {\cal H}_{bc}
\fe
annihilated by $b_0^-$ and $L_0^-$.
In particular, the latter admits a nontrivial BRST representative of the form
\ie
c\wt c\, {\bf \varphi}^{\mathbb{A}_\lambda} \Upsilon_{1-h, 1-\wt h}
\fe
{\it if} $\mathbb{A}_\lambda$ admits a primary ${\bf \varphi}^{\mathbb{A}_\lambda}$ of weight $(h, \wt h)$. There should be a corresponding $Q_{\Psi_\lambda}$-cohomology represented by a solution to (\ref{linearhh}) of the form (\ref{phihh}) with ghost number 2.

On the other hand, for a compact CFT $\mathbb{A}_\lambda$, at generic $(h, \wt h)$ there is no nontrivial BRST-cohomology in (\ref{alambdhh}), and there should be no corresponding $Q_{\Psi_\lambda}$-cohomology of the form (\ref{phihh}). Therefore, by seeking nontrivial solutions to (\ref{linearhh}), which may be constructed in perturbation theory, we will be able to determine the possible weights $(h, \wt h)$ of primaries of $\mathbb{A}_\lambda$. The extraction of further CFT data such as correlation functions from the SFT approach will be discussed in section \ref{sec:corramp}.

To proceed, we can expand $\Phi_{h,\wt h}$ on a basis of Virasoro descendants of the mock primary as
\ie\label{phihha}
\Phi_{h(\lambda), \wt h(\lambda)} =  \sum_{\underline N,\underline {\wt N}} \Phi^{\underline N,\underline {\wt N}}(\lambda) {\cal L}^{\mathbb M}_{-\underline{N}}\wt {\cal L}^{\mathbb M}_{-\underline{\wt N}} \Upsilon_{1-h(\lambda), 1-\widetilde h(\lambda)},
\fe
where ${\underline N}\equiv\{n_1,\cdots,n_k\}$ is a sequence of non-negative integers in descending order, ${\cal L}^{\mathbb M}_{-\underline{N}} \equiv L_{-n_1}^{\mathbb M}\cdots L_{-n_k}^{\mathbb M}$ is a complete basis of descending operators of the $c=26-c_{\mathbb A}$ holomorphic Virasoro algebra of the auxiliary CFT $\mathbb{M}$, and similarly for its anti-holomorphic counterpart $\wt {\cal L}^{\mathbb M}_{-\underline{\wt N}}$.\footnote{For simplicity, we will assume that the primary weights are generic so that there are no null descendants.} We will also write $|N|\equiv n_1+\cdots +n_k$ for the level of the Virasoro descendant.

Note that in (\ref{phihha}) we have explicitly indicated the dependence of the weights $(h(\lambda), \wt h(\lambda))$ on the deformation parameter $\lambda$. Generally, we expect the spin $s=h(\lambda)-\wt h(\lambda)$ to be independent of $\lambda$, whereas the scaling dimension $\Delta(\lambda) = h(\lambda) + \wt h(\lambda)$ is to be determined by solving the linearized SFT equation (\ref{linearhh}). $\Phi^{\underline N,\underline {\wt N}}(\lambda)\in {\cal H}_{\mathbb A}\otimes {\cal H}_{bc}$ are subject to the initial condition at $\lambda=0$,
\ie{}
& \Phi^{\underline N,\underline {\wt N}}(0)= \left\{ \begin{array}{ll} c\wt c \varphi,~~~& |N|=|\wt N|=0 \\ 0,~~~ & |N|+|\wt N|\geq 1 \end{array} \right.
\fe
where $\varphi$ is a primary of the CFT $\mathbb{A}$ with weight $(h_\varphi, \wt h_\varphi) = (h(0), \wt h(0))$. Generally, when there are multiple primaries of $\mathbb{A}$ of the same weight, an operator-mixing problem must be solved to determine $\varphi$ as a suitable linear combination of a basis of degenerate primaries with $\lambda$-dependent coefficients, as is standard in degenerate perturbation theory. For simplicity of discussion, we will proceed by assuming that such degeneracies are absent.

There are two equivalent methods of carrying out the perturbation analysis of the fluctuation spectrum. The first method is to directly expand $\Phi_{h(\lambda), \wt h(\lambda)}$ with respect to $\lambda$. This requires formally taking derivatives with respect to the weight of the mock primary and would necessitate working in a larger formal space of states. The second method, which is more convenient for practical computations, is to first eliminate the mock primaries by reducing the linearized SFT equation to a set of equations for $\Phi^{\underline N,\underline {\wt N}}(\lambda)$ and $\Delta(\lambda)$, which can then be solved perturbatively in $\lambda$. We now demonstrate the latter approach at the leading nontrivial order.

Writing
\ie
\Phi^{\underline N,\underline {\wt N}}(\lambda) = c\wt c \varphi \,\delta_{|N|+|\wt N|,0} + \sum_{n=1}^\infty \lambda^n \Phi^{\underline N,\underline {\wt N}}_n,
\fe
but keeping general $\lambda$-dependence in $(h(\lambda), \wt h(\lambda))$, we can expand the linearized SFT equation (\ref{linearhh}) in the background (\ref{marginalexpan}) as
\ie\label{firstorderphi}
Q_B c\wt c \varphi \Upsilon  + \lambda Q_B \sum_{\underline N,\underline {\wt N}} \Phi_1^{\underline N,\underline {\wt N}} {\cal L}^{\mathbb M}_{-\underline{N}}\wt {\cal L}^{\mathbb M}_{-\underline{\wt N}} \Upsilon + \lambda \left[ \Psi_1 \otimes c\wt c \varphi \Upsilon \right] = {\cal O}(\lambda^2),
\fe
where we have used the abbreviated notation $\Upsilon$ for $\Upsilon_{1-h(\lambda), 1-\wt (\lambda)}$. The notation ``${\cal O}(\lambda^2)$'' on the RHS means states built out of Virasoro descendants of $\Upsilon$ with ${\cal O}(\lambda^2)$ coefficients. Imposing the Siegel gauge condition $b_0^+ \Phi^{\underline N,\underline {\wt N}}_n=0$, 
we can act $b_0^+$ on (\ref{firstorderphi}) to obtain
\ie\label{deltaeqnt}
& (\Delta(0) - \Delta(\lambda)) c\wt c \varphi \Upsilon + \lambda \sum_{\underline N,\underline {\wt N}} ((L_0^{\mathbb{A}} )^++(L_0^{\rm gh})^+ +|N|+|\wt N|+2-\Delta(\lambda)) \Phi_1^{\underline N,\underline {\wt N}} {\cal L}^{\mathbb M}_{-\underline{N}}\wt {\cal L}^{\mathbb M}_{-\underline{\wt N}} \Upsilon
\\
&+ \lambda b_0^+ \left[ \Psi_1 \otimes c\wt c \varphi \Upsilon \right] = {\cal O}(\lambda^2).
\fe
Using $\Psi_1 = c\wt c{\cal O}$ (\ref{psiona}), and working with the flat-vertex frame string field (as defined in section \ref{sec:flatvert}), we can write the $\mathbb{M}$-level 0 component of (\ref{deltaeqnt}) as
\ie{}
& (\Delta(0) - \Delta(\lambda)) c\wt c \varphi + \lambda (L_0^+ +2 - \Delta(\lambda)) \Phi_1^{\{\},\{\}} 
\\
&+ \lambda b_0^+b_0^- r_0^{-L_0^+ -2 +\Delta(\lambda)} \left(c\wt c{\cal O}(-z_0) c\wt c \varphi(z_0) \right)|_{L_0^-=s} = {\cal O}(\lambda^2),
\fe
where in the third term, the restriction to the spin $L_0^- = s$ component comes from stripping off $\Upsilon$ followed by the spin projection $\mathbb{P}^-$.
Further projecting onto the ghost sector ground state, we have
\ie\label{reddphieq}
(\Delta(0) - \Delta(\lambda)) \varphi + \lambda (L_0^+ - \Delta(\lambda)) \Phi_1^{\{\},\{\}}|_{c\wt c} + 2 \lambda r_0^{\Delta(\lambda)-L_0^+} |2z_0|^2 \left({\cal O}(-z_0) \varphi(z_0) \right) |_{L_0^-=s} = {\cal O}(\lambda^2)
\fe
as an equation in ${\cal H}_{\mathbb A}$. Due to the freedom of overall rescaling of the fluctuation string field $\Phi$, we can assume without loss of generality that $\Phi_1^{\{\},\{\}}|_{c\wt c}$ is orthogonal to $\varphi$. Further assuming that $\varphi$ is the only primary in ${\cal H}_{\mathbb A}$ of weight $(h_\varphi, \wt h_\varphi)$, $h_\varphi - \wt h_\varphi =s$, taking the overlap of (\ref{reddphieq}) with $\varphi$ then determines\footnote{Here $\bar\varphi$ is the Hermitian conjugate of $\varphi$, and we have normalized $\langle\varphi \bar\varphi\rangle = 1$, assuming that the CFT $\mathbb{A}$ is compact.}
\ie
\Delta(\lambda) = \Delta(0) + 2\lambda C_{{\cal O}\varphi\bar\varphi} + {\cal O}(\lambda^2),
\fe
in agreement with the result of first order conformal perturbation theory.

\subsection{Decoupling the linear dilaton sector}
\label{sec:decoupling}

To study deformations of CFTs that change the central charge, as discussed in section \ref{sec:ccshift}, we replace $\mathbb{A}$ by its tensor product with a linear dilaton CFT $\mathbb{D}_\beta$. To ensure that the string field solution $\Psi_\lambda$ corresponds to a deformed worldsheet CFT of the factorized form (\ref{admfact}), a criterion (\ref{jpsicri}) for the decoupling of the linear dilaton sector was introduced, which we now analyze in more detail.

The ghost dilaton ${\cal K}_0$ (\ref{jkzero}) can be written as
\ie
{\cal K}_0 = Q_B \varkappa,
\fe
where
\ie
\varkappa \equiv {1\over 2} (\partial c - \bar\partial \wt c) = c_0^-\cdot 1,
\fe
is {\it not} a closed string field since it is not annihilated by $b_0^-$. This can be generalized to an expression for the string field that describes a constant dilaton shift around a nontrivial background solution $\Psi$ \cite{Bergman:1994qq}, of the form
\ie
{\cal K}_\Psi = Q_B \varkappa - \Psi + [\Psi\otimes \varkappa] + \cdots,
\fe
where $\cdots$ stands for higher order terms in $\Psi$. The full expression of ${\cal K}_\Psi$ is given by the Hamiltonian vector field defined by (4.8) and (8.3) of \cite{Bergman:1994qq}, shifted by $-\Psi$ due to our normalization of the string field (such that the string coupling drops out of the equation of motion), which can be further transformed into the flat-vertex frame according to the prescription of section \ref{sec:flatvert}. The decoupling criterion (\ref{jpsicri}) seeks to trivialize ${\cal K}_\Psi$ through a ghost number 1 string field ${\cal J}_\Psi$ that corresponds to the shift symmetry in the linear dilaton direction.

In the example of SFT solution describing a short RG flow of section \ref{sec:ccthird}, we have seen that a nontrivial dilaton profile is turned on at order $\lambda^3$, and therefore at the next order in $\lambda$ the SFT solution will necessarily involve products of non-identity primaries from both $\mathbb{A}$ and $\mathbb{D}_\beta$, potentially leading to a cohomological obstruction for solving (\ref{jpsicri}), namely
\ie
Q_\Psi {\cal J}_\Psi + \B' {\cal K}_\Psi = R_\Psi
\fe
is an order $\lambda^4$ string field that represents a nontrivial $Q_\Psi$-cohomology. We can remove this obstruction by correcting the SFT solution $\Psi$ to $\Psi + \Delta \Psi$, where $\Delta\Psi$ satisfies
\ie{}
\left[ \Delta\Psi\otimes {\cal J}_0 \right] + \B' (-\Delta\Psi + [\Delta\Psi\otimes\varkappa]) = R_\Psi +  ({\rm higher~order}).
\fe
Such a procedure allows for constructing a solution that satisfies the decoupling criterion (\ref{jpsicri}) order by order in $\lambda$, as desired.

\subsection{CFT correlators from string amplitudes}
\label{sec:corramp}

The correlation functions of the deformed CFT $\mathbb{A}_\lambda$ can be extracted from the on-shell string amplitudes in the background (\ref{newamcft}), which are reproduced in the SFT as the tree-level amplitudes of on-shell fluctuations around the solution $\Psi_\lambda$. For this purpose, it is necessary to consider the SFT action functional, given by \cite{Zwiebach:1992ie}
\ie\label{sftaction}
S[\Psi] = {1\over 2} \langle \Psi| c_0^- Q_B |\Psi \rangle + \sum_{n=3}^\infty {1\over n!} \{\Psi^{\otimes n}\},
\fe
where $\langle\cdot|$ stands for BPZ conjugate, and $\{\cdots\}$ stands for {\it symmetric} string vertices that obey the BV master equation (\ref{bvmaster}). The variational principle of (\ref{bvmaster}) leads to the equation of motion (\ref{sfteom}).

In the Siegel gauge $b_0^+\Psi=0$, using $\{Q_B, b_0^+\}=L_0^+$ one can write the kinetic term as
\ie
{1\over 2} \langle \Psi| c_0^- Q_B |\Psi \rangle = {1\over 2} \langle \Psi| c_0^- c_0^+ L_0^+  |\Psi \rangle,
\fe 
giving rise to a propagator proportional to $1/L_0^+$.
The $n$-point tree-amplitude of on-shell linearized fluctuations $\Phi_1,\cdots,\Phi_n\in\hat{\cal H}$ that obey $b_0^+\Phi_i=Q_B\Phi_i=0$ can be turned into a single integral over ${\cal M}_{0,n}$ in the familiar form
\ie\label{ampsimp}
{\cal A}_{\mathbb{A}}(\Phi_1, \ldots, \Phi_n) = \frac{1}{(- 2\pi i)^{n -3}} \int_{{\cal M}_{0, n}}\left \langle \prod_{k = 1}^{2n- 6 } \mc B_{k} dt^k \prod_{i = 1}^n \Phi_i \right\rangle_{\mathbb{A}\otimes \mathbb{M}\otimes (bc)}.
\fe
In particular, if we take
\ie
\Phi_i = c\wt c \varphi_i \Upsilon_{1-h_i, 1- \wt h_i},
\fe
where $\varphi_i$ is a primary of $\mathbb{A}$ of weight $(h_i, \wt h_i)$, the integrand is determined via Virasoro Ward identities by the $n$-point function $\langle \varphi_1\cdots\varphi_n\rangle_{\mathbb A}$ in the CFT $\mathbb{A}$ and the $n$-point function of mock primaries
\ie\label{testfunc}
f(z_1,\cdots,z_n) \equiv \left\langle \prod_{i=1}^n \Upsilon_{1-h_i, 1-\wt h_i}(z_i) \right\rangle_{\mathbb M}
\fe
in the auxiliary CFT $\mathbb M$.

The SFT amplitude prescription can be straightforwardly extended to the nontrivial background solution $\Psi_\lambda$, via the action
\ie
S_{\Psi_\lambda}[\Psi'] &\equiv S[\Psi_\lambda + \Psi'] - S[\Psi_\lambda]
\\
&=  {1\over 2} \langle \Psi'| c_0^- Q_{\Psi_\lambda} |\Psi' \rangle + \sum_{m\geq 0, n\geq 3} {1\over m! n!} \{\Psi_\lambda^{\otimes m}\otimes \Psi'^{\otimes n}\}.
\fe
The $n$-point tree amplitudes of the on-shell fluctuations $\Phi_i\in\hat{\cal H}_{h_i,\wt h_i}$, $i=1,\cdots,n$, that obey $b_0^+\Phi_i=Q_{\Psi_\lambda}\Phi_i=0$, will be denoted\footnote{See \cite{Zwiebach:1992ie, Sen:2019jpm} for a detailed description of the Feynman rules based on the SFT action.}
\ie\label{ampex}
{\cal A}_{\mathbb A}(\Phi_1,\cdots,\Phi_n | \Psi_\lambda).
\fe
Note that (\ref{ampex}) depends on the auxiliary CFT data only through the $n$-function of mock primaries $f(z_1,\cdots, z_n)$ (\ref{testfunc}), as $\Psi_\lambda$ does not involve non-identity mock primaries.

Suppose the $Q_{\Psi_\lambda}$-cohomology of $\Phi_i$ corresponds to the on-shell vertex operator
\ie
\Phi_i^{\mathbb{A}_\lambda} = c\wt c \varphi_i^{\mathbb{A}_\lambda} \Upsilon_{1-h_i,1-\wt h_i}
\fe
in the deformed worldsheet CFT (\ref{newamcft}) in the manner described in section \ref{sec:spec}, then the amplitude (\ref{ampex}) should admit an equivalent expression
\ie\label{ampdefo}
{\cal A}_{\mathbb{A}_\lambda}(\Phi_1^{\mathbb{A}_\lambda} ,\cdots,\Phi_n^{\mathbb{A}_\lambda} )
\fe
defined as in (\ref{ampsimp}) except that the CFT $\mathbb{A}$ is replaced by $\mathbb{A}_\lambda$. In particular, (\ref{ampdefo}) is expressed as a integral over ${\cal M}_{0,n}$, whose integrand is determined via Virasoro Ward identities by the $n$-point function $\langle\varphi_1\cdots\varphi_n\rangle_{\mathbb{A}_\lambda}$ in the deformed CFT $\mathbb{A}_\lambda$, and the {\it same} $n$-point function of mock primaries $f(z_1,\cdots, z_n)$ as appearing in (\ref{ampex}). 

Equating (\ref{ampdefo}) with (\ref{ampex}) and considering functional variations with respect to $f(z_1,\cdots, z_n)$ then in principle determines the correlation functions of $\mathbb{A}_\lambda$ in terms of the CFT data of $\mathbb{A}$. In particular, the structure constants of $\mathbb{A}_\lambda$ are given by\footnote{Here it is important that one works with symmetric string vertices. A solution of the deformed string field equation (\ref{deformsfteom}) defined via the asymmetric flat vertices of section \ref{sec:flatvert} would be related to $\Psi_\lambda$ appearing in (\ref{cijka}) by a field redefinition obtained by integrating (\ref{dphis}).}
\ie\label{cijka}
C_{ijk}^{\mathbb{A}_\lambda} &= {\cal A}_{\mathbb A}(\Phi_i,\Phi_j,\Phi_k | \Psi_\lambda)
\\
&= \sum_{m=0}^\infty {1\over m!} \{ \Psi_\lambda^{\otimes m} \otimes \Phi_i\otimes \Phi_j\otimes\Phi_k \} ,
\fe
where the 3-point function coefficient of mock primaries appearing on the RHS is normalized to 1. This formula amounts to a precise all-order implementation of renormalized conformal perturbation theory. The background solution $\Psi_\lambda$ plays the role of a Wilsonian effective action at the RG fixed point. The choice of string vertices, subject to the geometric master equation (\ref{geometricmeq}), is analogous to a renormalization scheme. The dependence on the choice of string vertices can be absorbed by a string field redefinition, and drops out of the on-shell amplitude. The latter relies crucially on the $Q_{\Psi_\lambda}$-closure of $\Phi_i, \Phi_j, \Phi_k$. Furthermore, the SFT amplitude on the RHS of (\ref{cijka}) is manifestly finite, where the moduli integrations appearing in the string vertices by construction avoid the boundary of the moduli space where operators would collide.

\section{Application I: the Horowitz-Polchinski solution}
\label{sec:hp}

Consider the CFT $\mathbb{A}$ given by $d$ noncompact free bosons $\vec Z = (Z^1,\cdots, Z^d)$ together with a compact boson $X$ at radius $R={1\over \sqrt{2}}$ (i.e. half of the self-dual radius\footnote{As in section \ref{sec:ccshift}, we work in $\A'=2$ convention.}), and a string field of the form 
\ie\label{hpansatz}
&\Psi_\lambda = W_\lambda + (1-\mathbb{P}_\lambda^+) \Psi_\lambda,
\\
&W_\lambda = \lambda  \Big[ c\wt c f(\sqrt{\lambda} \vec Z) \cos(\sqrt{2} X) + c\wt c g(\sqrt{\lambda} \vec Z) \partial X \bar\partial X 
%\\&~~~~~~~~~~ +  \phi(\sqrt{\lambda} \vec Z)  \left( c\wt c \partial \vec Z\cdot \bar\partial\vec Z - {1\over 2}(c\partial^2c - \wt c\bar\partial^2\wt c)  \right) 
\Big] + {\cal O}(\lambda^3),
\fe
where $\mathbb{P}_\lambda^+$ stands for the projector onto states with $|L_0^+|\lesssim {\cal O}(\lambda)$, and $f,g$ are smooth and bounded spherically symmetric functions on $\mathbb{R}^d$. Importantly, we will keep track of the $\lambda$ dependence of $f(\sqrt{\lambda}\vec Z)$ and $g(\sqrt{\lambda}\vec Z)$ in the region of large $|\vec Z|$ in the field target space, while omitting higher order corrections that are suppressed by powers of $\lambda$ globally in the target space.

As in (\ref{weqn}), the SFT equation for $\Psi_\lambda$ can be brought to the form
\ie
Q_B W_\lambda + {1\over 2} \mathbb{P}_\lambda^+ [W_\lambda^{\otimes 2}] = {\cal O}(\lambda^3).
\fe
The ansatz (\ref{hpansatz}) gives
\ie{}
& Q_B W_\lambda = \lambda^2 c_0^+ \Big[  c\wt c (-\nabla^2 f)(\sqrt{\lambda} \vec Z) \cos(\sqrt{2} X) +  c\wt c (-\nabla^2 g)(\sqrt{\lambda} \vec Z) \partial X \bar\partial X 
%\\ &~~~~~~~~~~~ + (-\nabla^2 \phi)(\sqrt{\lambda} \vec Z) \left(  c\wt c \partial \vec Z\cdot \bar\partial\vec Z -{1\over 2} (c\partial^2c - \wt c\bar\partial^2\wt c) \right)  
\Big] + {\cal O}(\lambda^3),
\fe
and with the 2-string bracket in the flat-vertex frame,
\ie{}
\mathbb{P}_\lambda^+ [W_\lambda^{\otimes 2}] &= b_0^-\mathbb{P}^- r_0^{-L_0^+} \mathbb{P}_\lambda^+ (W_\lambda(-z_0) W_\lambda(z_0))
\\
&= \lambda^2 2c_0^+c\wt c  \left[ -2 f(\sqrt{\lambda} \vec Z)  g(\sqrt{\lambda} \vec Z)  \cos(\sqrt{2}X) - (f(\sqrt{\lambda} \vec Z))^2 \partial X \bar\partial X \right] + {\cal O}(\lambda^3).
\fe
Note that in particular, various terms in the $W_\lambda(-z_0) W_\lambda(z_0)$ OPE that involve the gradient of $f$ and $g$ are of order $\lambda^3$.

Therefore, the SFT equation is satisfied up to ${\cal O}(\lambda^3)$ terms provided
\ie\label{fgeqn}
& \nabla^2 f + 2 f g =0,
\\
& \nabla^2 g + f^2=0.
\fe
These are precisely equations (2.7), (2.8) of \cite{Chen:2021dsw} up to a simple rescaling and constant shift. For $d=3,4,5$, a smooth, bounded, spherically symmetric solution to (\ref{fgeqn}) exists, with the asymptotic behavior
\ie\label{fgbdry}
f(\infty)=0, ~~~~ g(\infty) \equiv g_\infty <0.
\fe
The string field (\ref{hpansatz}) with $f,g$ satisfying (\ref{fgeqn}) and (\ref{fgbdry}) describes none other than the T-dual of the Horowitz-Polchinski ``string star'' \cite{Horowitz:1997jc}. 

In contrast to the analysis of \cite{Horowitz:1997jc, Chen:2021dsw} which is based on massless effective field theory in $d$ dimensions, here we have given a first principle construction of the corresponding solution $\Psi_\lambda$ in the classical bosonic closed SFT in a self-consistent weak field approximation, which can nonetheless be systematically improved by computing higher order corrections in $\lambda$.\footnote{The effective field theory with quartic interaction has been analyzed in \cite{Balthazar:2022hno}.} From the worldsheet conformal perturbation theory perspective, the smoothness of the functions $f$ and $g$ at the origin in $\mathbb{R}^d$ may come as a surprise; it signifies that the string field captures renormalized rather than bare couplings.

\section{Application II: tachyon-dilaton eschatology via Runkel-Watts walls}
\label{sec:escha}

Let the CFT $\mathbb{A}$ be the $c=1$ free {\it noncompact} boson, and consider its deformation by the weight $(1,1)$ primary $\cos(\sqrt 2 X)$. It is well known that the analogous deformation in a compact boson theory would not be exactly marginal \cite{Chaudhuri:1988qb}, as a nonzero beta function would be generated at second order in conformal perturbation theory. In the noncompact setting, however, it is possible to correct the deformation at each order in the coupling to cancel the beta function at the price of breaking discrete shift symmetry in $X$, so as to preserve marginality to all orders in perturbation theory. We will investigate such a perturbatively marginal deformation using the SFT approach by embedding $\mathbb{A}$ into the worldsheet CFT along with a linear dilaton sector $\mathbb{D}_\beta$, as introduced in section \ref{sec:ccshift}. From the SFT perspective, the deformation amounts to turning on a closed string tachyon profile that varies in one spatial dimension, and we will analyze the fate of such a tachyon solution with nonlinear effects taken into account.

\subsection{The perturbative SFT solution up to second order}

We begin by seeking a SFT solution $\Psi_\lambda$ of the perturbative form (\ref{marginalexpan}) with the first order string field
\ie
\Psi_1 = \lambda c \wt c \cos (\sqrt 2 X). 
\fe
Following the strategy of section \ref{sec:marginal}, we can write the second order string field solution as
\ie
\Psi_2 = W_2 - {1\over 2} {b_0^+\over L_0^+} (1-\mathbb{P}^+) [\Psi_1^{\otimes 2}],
\fe
where $W_2$ is a weight $(0,0)$ string field that obeys
\ie\label{qbetwocos}
Q_B W_2 = -{1\over 2} \mathbb{P}^+ [\Psi_1^{\otimes 2}] = c_0^+ c\wt c \partial X \bar\partial X.
\fe
In the absence of the linear dilaton sector $\mathbb{D}_\B$, the RHS of (\ref{qbetwocos}) would in fact represent a nontrivial $Q_B$-cohomology class of ghost number 3 that obstructs the second order SFT solution, corresponding to a central charge shift of the deformed CFT $\mathbb{A}_\lambda$ at order $\lambda^2$. This is mostly easily understood by rewriting (\ref{qbetwocos}) as
\ie\label{qbewaf}
Q_B W_2 = c_0^+ D_X + {1\over 2} c_0^+(c\partial^2 c -\wt c \bar\partial^2\wt c),
\fe
where $D_X = c\wt c \partial X \bar\partial X - {1\over 2} (c\partial^2 c -\wt c \bar\partial^2\wt c)$ is the zero momentum limit of the vertex operator that corresponds to a dilaton profile in the $X$ direction \cite{Bergman:1994qq}. Using the formula
\ie
Q_B (f(X)D_X) = - c_0^+ f''(X) D_X,
\fe
we can solve
\ie\label{wtwodit}
W_2 = -{1\over 2} X^2 D_X + {1\over 2\B} \Theta,
\fe
where $\Theta$ is the string field that describes a shift of the background charge of $\mathbb{D}_\beta$ as defined in (\ref{dexpr}). In the notation of (\ref{antighosd}), (\ref{carel}), we have the anti-ghost dilaton coefficient $a={1\over 2}\lambda^2$ from (\ref{qbewaf}), and thus a second order shift of the central charge of the deformed CFT $\mathbb{A}_\lambda$,
\ie\label{cadefa}
c_{\mathbb{A}_\lambda} = 1 - 6\lambda^2 + {\cal O}(\lambda^3).
\fe
The decoupling of the extraneous $Y$-linear dilaton sector is manifest in (\ref{wtwodit}). Starting at order $\lambda^3$, it will be necessary to constrain the solution so as to maintain the decoupling as discussed in section \ref{sec:decoupling}.

\subsection{Extending the solution to large field region}

We observe that (\ref{wtwodit}) describes a dilaton profile in $X$ direction of the form $\phi \sim {1\over 2} \lambda^2 X^2$, which is slowly varying at small $\lambda$ but grows with $X$. By analogy with the Horowitz-Polchinski solution considered in section \ref{sec:hp}, we postulate the following ansatz that capture the slow-varying components of the string field at small $\lambda$,
\ie\label{wlamansat}
W_\lambda \equiv \mathbb{P}_{\lambda^2}^+\Psi_\lambda &= {\lambda\over 2} c\wt c \left(f(\lambda X) e^{i\sqrt{2} X} + \bar f(\lambda X) e^{-i\sqrt{2} X} \right)+ \phi(\lambda X) {\cal K}_0
\\
& ~~~ + \lambda A(\lambda X) c_0^+ (c\partial X - \wt c \bar\partial X) + c\wt c \, h(\lambda X)\partial X \bar\partial X
 + {\lambda^2 \over 2\beta} \Theta + {\cal O}(\lambda^3),
\fe
where $\mathbb{P}_{\lambda^2}^+$ stands for the projection onto states with $|L_0^+|\lesssim {\cal O}(\lambda^2)$, ${\cal K}_0$ is the ghost dilaton defined in (\ref{jkzero}), $\Theta$ is defined in (\ref{dexpr}), and $f(x), \phi(x), A(x), h(x)$ are smooth functions of $x$.\footnote{The functions $f(x), \phi(x), A(x), h(x)$ themselves may depend on $\lambda$ analytically, but this won't affect the analysis of the SFT solution at the order we consider below.}

On the RHS of (\ref{wlamansat}) we have included all possible ghost number 2 string fields with $|L_0^+|\lesssim {\cal O}(\lambda^2)$ that are even with respect to worldsheet parity\footnote{Note that by convention $c\wt c$ is parity even whereas the identity operator is parity odd.} while maintaining the decoupling of the $Y$-linear dilaton sector, the only assumption being the $\lambda$-scaling which will be justified by solving the SFT equation below. The functions $f, \phi, A, h$ are not necessarily bounded, and the separation of these slow-varying fields from the higher order corrections in $\lambda$ is meaningful for the range of $X$ of order $\sim 1/ \lambda$. Note that we have not assumed the Siegel gauge condition in (\ref{wlamansat}); as already seen in section \ref{sec:ccshift}, this will allow for the flexibility of making other more convenient gauge choices in computing the solution.

A priori, the appearance of $\phi(\lambda X)$ and $h(\lambda X)$ in (\ref{wlamansat}) which are of order one would mean that all higher-point string vertices would contribute to the SFT equation that governs $W_\lambda$. However, this complication can be evaded as follows. Firstly, we will work in a gauge in which
\ie
h(\lambda X) = 0.
\fe
Note that by doing so, we no longer have the freedom of gauging away $A(\lambda X)$ in (\ref{wlamansat}). Secondly, we will work in the flat-vertex frame defined in section \ref{sec:flatvert}, in which the flat string bracket can never reduce the number of derivatives in a $\partial^2c$ appearing in a string field, and consequently the string bracket contributing to the equation of motion for $W_\lambda$ can never involve more than one insertion of $\phi(\lambda X){\cal K}_0$. This is a major simplification that allows us to write down the full equation for (\ref{wlamansat}) in the flat-vertex frame using only 2-string brackets.

Splitting the $X$-dependent part of (\ref{wlamansat}) into the ``tachyon component''
\ie\label{wtdef}
W_T =  {\lambda\over 2} c\wt c \left(f(\lambda X) e^{i\sqrt{2} X} + \bar f(\lambda X) e^{-i\sqrt{2} X} \right)
\fe
and the ``dilaton component''
\ie\label{wddef}
W_D =  \phi(\lambda X) {\cal K}_0 + \lambda A(\lambda X) c_0^+ (c\partial X - \wt c \bar\partial X) + c\wt c\, h(\lambda X)\partial X \bar\partial X,
\fe
we can reduce the SFT equation, in the form of (\ref{weqn}), to
\ie\label{qwww}
& Q_B W_T = - \mathbb{P}_{\lambda^2}^+ [W_T\otimes W_D] + {\cal O}(\lambda^3),
\\
& Q_B W_D = -{1\over 2} \mathbb{P}_{\lambda^2}^+ [W_T^{\otimes 2}]-{1\over 2}\mathbb{P}_{\lambda^2}^+ [W_D^{\otimes 2}] - \lambda^2 c_0^+ {\cal K}_0 + {\cal O}(\lambda^3).
\fe

On the LHS of (\ref{qwww}), the BRST variations of $W_T$ and $W_D$ are computed from (\ref{wtdef}) and (\ref{wddef}), up to higher order terms in $\lambda$, as
\ie{}
& Q_B W_T = \sqrt{2}\lambda^2 c_0^+ c\wt c (-i f'(\lambda X) e^{i\sqrt{2} X} + i \bar f'(\lambda X) e^{-i\sqrt{2} X}) + {\cal O}(\lambda^3),
\\
& Q_B W_D =  -\lambda^2 (\phi''(\lambda X) + A'(\lambda X))c_0^+{\cal K}_0 + 
\lambda (\phi'(\lambda X) - A(\lambda X)) (c\partial X + \wt c \bar\partial X) {\cal K}_0
\\
&~~~~~~~~~~~~ + 2\lambda^2 A'(\lambda X) c_0^+ c\wt c \partial X \bar\partial X + {\cal O}(\lambda^3).
\fe
The projected 2-string brackets appearing on the RHS of (\ref{qwww}) are computed, in the flat-vertex frame,\footnote{Note that with a non-flat string vertex, there would be additional terms on the RHS, particularly those that involve $\phi(\lambda X)$.} as
\ie{}
& - \mathbb{P}_{\lambda^2}^+ [W_T\otimes W_D]  = \sqrt{2} \lambda^2 c_0^+ c\wt c (-i f(\lambda X) A(\lambda X) e^{i\sqrt{2} X} + i \bar f(\lambda X) A(\lambda X) e^{i\sqrt{2} X} ) + {\cal O}(\lambda^3),
\\
& -{1\over 2} \mathbb{P}_{\lambda^2}^+ [W_T^{\otimes 2}] = \lambda^2 |f(\lambda X)|^2 c_0^+ c \wt c \partial X \bar\partial X+ {\cal O}(\lambda^3),
\\
& -{1\over 2}\mathbb{P}_{\lambda^2}^+ [W_D^{\otimes 2}] = - 2 \lambda^2 (\phi'(\lambda X))^2 c_0^+ {\cal K}_0+ {\cal O}(\lambda^3).
\fe
Up to order $\lambda^2$, (\ref{qwww}) is thus reduced to the differential equations 
\ie\label{afdiffeq}
& A(x) = \phi'(x),
\\
& f'(x) = f(x) A(x),
\\
& A'(x) = {1\over 2} |f(x)|^2,
\\
& \phi''(x) + A'(x) = 2 (\phi'(x))^2 + 1.
\fe
After eliminating $A(x)$ with the first equation of (\ref{afdiffeq}), we can solve the second equation of (\ref{afdiffeq}) with
\ie
f(x) = e^{\phi(x)-\phi_0},
\fe
where $\phi_0 = \phi(0)$ by the normalization convention $f(0)=1$. The remaining equations of (\ref{afdiffeq}) can now be expressed as
\ie
\phi''(x)  = (\phi'(x))^2 + {1\over 2} = {1\over 2} e^{2\phi(x)-2\phi_0},
\fe
where the first equality is in fact redundant, and we can solve the first order differential equation for $\phi$ to give
\ie
\phi(x) = \phi_0 - \log \big( \cos {x\over \sqrt{2}} \big).
\fe
The resulting solution of the projected string field (\ref{wlamansat}), in the $h=0$ gauge and in the flat-vertex frame, is
\ie\label{wlambsol}
W_\lambda = \lambda c\wt c\, {\cos(\sqrt{2} X) \over \cos({\lambda X\over \sqrt{2}})} + \Big(\phi_0 - \log \big( \!\cos {\lambda X\over \sqrt{2}} \big) \Big) {\cal K}_0 + {\lambda\over \sqrt{2}} \tan\!\big({\lambda X\over \sqrt{2}}\big) c_0^+ (c\partial X - \wt c \bar\partial X)  +  {\lambda^2 \over 2\beta} \Theta + {\cal O}(\lambda^3).
\fe
From the target space perspective, this describes a string background with a weak tachyon field of spatial momentum $p\simeq \pm {1\over \sqrt{2}}$ whose amplitude $\lambda f(\lambda X)$ is further modulated over distance scale $1/\lambda$, and a dilaton profile $\phi(\lambda X)$ that grows with increasing $|X|$. Conspicuously, the dilaton diverges at 
\ie\label{xbdry}
X = \pm {\pi\over \sqrt{2} \lambda} + {\cal O}(\lambda^0),
\fe
and thus (\ref{wlambsol}) is expected to be valid only in the regime ${\pi\over \sqrt{2}\lambda}\lesssim X\lesssim {\pi\over \sqrt{2}\lambda}$, and describes a deformed background of finite space with walls at $|X|\approx {\pi\over \sqrt{2}\lambda}$.

\begin{figure}
	\def\svgwidth{1\linewidth}
	\centering{
		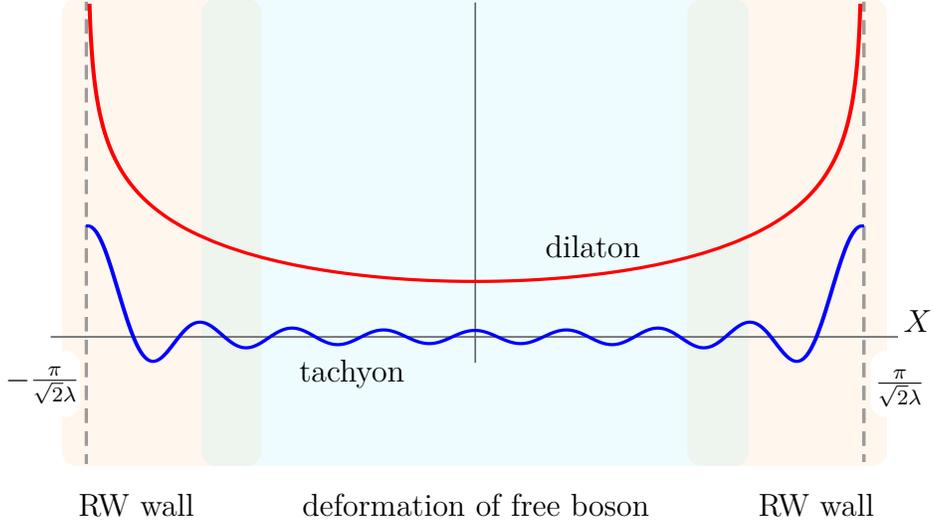	\caption{A schematic portrait of the tachyon and dilaton profile of the SFT solution \ref{wlambsol}. The perturbatively-marginal deformation of the free boson a priori describes the weak field region (shaded blue) of space which terminates at the Runkel-Watts walls.
			\label{fig:RWWall}
	}}
\end{figure}

\subsection{Runkel-Watts walls and the emergence of minimal model}

Following the strategy outlined in section \ref{sec:spec}, we can analyze the low-lying spectrum of the deformed CFT $\mathbb{A}_\lambda$ by considering a linearized string field fluctuation $\Phi$ over the background solution (\ref{wlambsol}) of the general form (\ref{phihh}), with
\ie\label{phians}
\Omega\equiv \mathbb{P}_{\lambda^2}^+\Phi = c\wt c \, \xi(\lambda X) \Upsilon + {\cal O}(\lambda^3),
\fe
where $\Upsilon$ is a mock primary of weight $(1-{\Delta\over 2}, 1-{\Delta\over 2})$ in the auxiliary CFT on the worldsheet, with 
\ie
\Delta = \lambda^2 \gamma +{\cal O}(\lambda^3) ,
\fe
and $\xi(\lambda X)$ is a wave function of wave length $\sim 1/\lambda$ to be determined. The linearized SFT equation can be put in the form
\ie\label{qomega}
Q_B \Omega + \mathbb{P}_{\lambda^2}^+ \left[W_\lambda \otimes \Omega \right]
+ {1\over 2}\mathbb{P}_{\lambda^2}^+ \left[W_\lambda^{\otimes 2}\otimes \Omega\right]^{\cal A} = {\cal O}(\lambda^3),
\fe
where we have defined the symmetric tri-linear map $[\cdot]^{\cal A}:\otimes \hat{\cal H}^{\otimes 3}\to \hat{\cal H}$ with
\ie
\left[\Psi^{\otimes 3}\right]^{\cal A} \equiv [\Psi^{\otimes 3}] + 3 \big[ {b_0^+\over L_0^+}(1-\mathbb{P}_{\lambda^2}^+) [\Psi^{\otimes 2}]\otimes \Psi \big]  .
\fe
Working with the flat string bracket which cannot reduce the number of derivatives on $\partial^2c$, we need not be concerned with the $\phi(\lambda X){\cal K}_0$ term in $W_\lambda$ (\ref{wlamansat}), and thus can treat $W_\lambda$ as of order $\lambda$ in (\ref{qomega}) and furthermore reduce (\ref{qomega}) to
\ie\label{linomeeq}
 Q_B \Omega + \mathbb{P}_{\lambda^2}^+ \left[W_D \otimes \Omega \right] + {1\over 2} \mathbb{P}_{\lambda^2}^+\left[W_T^{\otimes 2}\otimes \Omega\right]^{\cal A}  + {1\over 2} \mathbb{P}_{\lambda^2}^+ \left[W_D^{\otimes 2}\otimes \Omega \right]^{\cal A}  = {\cal O}(\lambda^3).
\fe 
The first two terms in (\ref{qomega}) are evaluated as
\ie\label{omsimp}
& Q_B \Omega = \lambda^2(- \xi''(\lambda X) - \gamma \xi(\lambda X)) c_0^+ c\wt c \Upsilon + {\cal O}(\lambda^3)
\\
& \mathbb{P}_{\lambda^2}^+ [W_D\otimes \Omega] = 2 \lambda^2 c_0^+ c \wt c A(\lambda X) \xi'(\lambda X) \Upsilon + {\cal O}(\lambda^3).
\fe
The remaining terms can be evaluated as in a naive computation of a 4-point on-shell amplitude, but with suitable divergences subtracted. Relying crucially on simplifications due to the flat string bracket, we find
that both $\mathbb{P}_{\lambda^2}^+\left[W_T^{\otimes 2}\otimes \Omega\right]^{\cal A}$ and $ \mathbb{P}_{\lambda^2}^+\left[W_D^{\otimes 2}\otimes \Omega\right]^{\cal A}$ vanish up to ${\cal O}(\lambda^3)$ terms.\footnote{$\mathbb{P}_{\lambda^2}^+\left[W_T^{\otimes 2}\otimes \Omega\right]^{\cal A}$ is computed by
\ie\label{pwwomcal}
{\lambda^2\over 2} {b_0^-\mathbb{P}^-\mathbb{P}_{\lambda^2}^+\over (-2\pi i)} \int'_{{\cal M}_{0,4}} dt\wedge d\bar t B_t B_{\bar t} {c\wt c(z_1) c\wt c(z_2) c\wt c(z_3)\over |z_{12}|^4 } |f|^2\xi \Upsilon = {\lambda^2\over 2\pi i} c_0^+ c\wt c |f|^2\xi  \Upsilon \int_{{\cal M}_{0,4}}'  \left| { z_{23} dz_1+ z_{31} dz_2 + z_{12} dz_3 \over z_{12}^2 }\right|^2
\fe
up to ${\cal O}(\lambda^3)$ terms, where $z_1, z_2, z_3$ are functions of the moduli $t,\bar t$ as defined in (\ref{coordmapexp}), and the integration is over the entire moduli space ${\cal M}_{0,4}$ but subtracting the divergent powers of $z_{12}$ as $z_{12}\to 0$ while fixing $z_3$ and ${z_1+z_2\over 2}$. Note that this manipulation is only valid for flat string brackets. Writing the integrand in the second line of (\ref{pwwomcal}) as
$\left| d \big({z_3 - {z_1+z_2\over 2} \over z_{12}} \big) \right|^2$,
one immediately sees that the result vanishes after subtracting divergences. $\mathbb{P}_{\lambda^2}^+\left[W_D^{\otimes 2}\otimes \Omega\right]^{\cal A}$ is computed by
\ie
2{\lambda^2}  {b_0^-\mathbb{P}^-\mathbb{P}_{\lambda^2}^+\over (-2\pi i)} \int'_{{\cal M}_{0,4}} dt\wedge d\bar t B_t B_{\bar t} { c_0^+ c(z_1)  c_0^+ c(z_2) c\wt c(z_3)\over z_{12}^2 }A^2\xi \Upsilon
\fe
up to ${\cal O}(\lambda^3)$ terms, which can only receive contributions from $(b_0^+)^{(z_i)} b_{-1}^{(z_j)}$ in $B_t B_{\bar t}$ (\ref{btexpr}). The latter give rise to an integrand proportional to $d \big({z_3 - {z_1+z_2\over 2} \over z_{12}} \big)\wedge d\log|{q_1\over q_2}|$, which again vanishes upon subtracting divergent powers of $z_{12}$.
} 
Combining this with (\ref{linomeeq}), (\ref{omsimp}), we obtain the differential equation 
\ie\label{xidiff}
\xi''(x) + \C \xi(x) - 2 A(x) \xi'(x) %+ \big( -{1\over 2} |f(x)|^2 + (A(x))^2\big) \xi(x) 
= 0.
\fe
Using $A(x) = \phi'(x)$ and ${1\over 2} |f(x)|^2 = \phi''(x)$ from (\ref{afdiffeq}), we see that (\ref{xidiff}) is equivalent to
\ie\label{xiphieq}
\left(\partial_x^2 + \C + {1\over 2} \right)\left(e^{-\phi(x)} \xi(x) \right) = 0.
\fe
The behavior $\phi(x)\to +\infty$ as $x\to \pm {\pi\over \sqrt{2}}$ suggests a Dirichlet type boundary condition for $e^{-\phi(x)} \xi(x)$, with $\xi(x)$ itself behaving regularly at the boundary. An a priori justification of the latter requires going beyond the slow-varying field approximation analyzed thus far. 

However, there is a natural interpretation of the boundary condition, as follows. In the $\lambda\to 0$ limit, the tachyon-dilaton profile of the background solution is negligible except near the pair of boundary walls $X\sim\pm {\pi\over \sqrt{2}\lambda}$ which are far separated. The central charge of the corresponding deformed CFT $\mathbb{A}_\lambda$ (\ref{cadefa}) approaches 1. We propose that the deformed CFT describing each of the near-boundary regions in the $\lambda \to 0$ limit is the Runkel-Watts theory \cite{Runkel:2001ng}, which is an irrational unitarity CFT of central charge 1, and thus the full space can be viewed as sandwiched between a pair of ``Runkel-Watts walls'' (Figure \ref{fig:RWWall}). 

The primaries $\phi_p$ of Runkel-Watts theory are labeled by a positive real parameter $p$ excluding positive integer values, with conformal weight $h_p = \wt h_p = {p^2\over 4}$.\footnote{The parameter $p$ of the primary $\phi_p$ is denoted $x$ in \cite{Runkel:2001ng}.} These are identified with the tachyon modes of momentum $p/\sqrt{2}$ that reflects off the wall, and missing primaries at positive integer $p$ are presumably due to resonance effects that result from interaction with the tachyon background. The structure constants $c(p_1, p_2, p_3)$ for a triple of primaries $\phi_{p_1}, \phi_{p_2}, \phi_{p_3}$ are given in \cite{Runkel:2001ng}. As explained in Appendix \ref{sec:runkelwatts}, a more natural basis of primaries is in fact $\phi_p' = (-)^{\lfloor p\rfloor} \phi_p$, where $\lfloor p\rfloor$ is the integer part of $p$. Upon extension to negative $p$ by defining $\phi_{-p}=\phi_p$, or equivalently $\phi'_{-p}=-\phi_p'$, the Fourier transform of $\phi'_p$ with respect to $p$, denoted by $\wt \phi_x$, has three-point function that behave in the large $x_i$ limit as\footnote{As the operators in question are dominated by primaries of weight close to zero, the (worldsheet) coordinate dependence of the three-function becomes negligible.}
\ie
\wt c(x_1, x_2, x_3) \sim{\mathrm const}\cdot { \delta(x_{12})\delta(x_{13}) \over x_1}.
\fe
This structure constant can be interpreted as the contact interaction of three tachyons at location $x$ ($>0$) in the target space, with a spatially varying (string) coupling that diverges as $x\to 0$, or equivalently a dilaton profile $\phi(x) \sim - \log x$. This is in agreement with the near-boundary behavior of the SFT solution (\ref{wlambsol}) at small $\lambda$. Furthermore, the wave function of $\phi_p'$ in $x$-space is proportional to $\sin (p x)$, which in particular vanishes at $x=0$. This wave function is to be identified with $e^{-\phi(x)} \xi(x)$ in (\ref{xiphieq}) up to a shift of $x$ by $\pi\over\sqrt{2}$, thereby justifying the proposal that  $e^{-\phi(x)} \xi(x)$ should obey Dirichlet boundary condition.

The solutions to (\ref{xiphieq}), subject to the boundary condition $e^{-\phi(x)}\xi(x)\to 0$ as $x\to \pm {\pi\over \sqrt{2}}$, are
\ie
\xi(x) = {1\over \cos{x\over \sqrt{2}}} \sin{\ell (x+{\pi\over \sqrt{2}})\over \sqrt{2}} ,~~~~\C = {1\over 2}(\ell^2-1),~~~ \ell= 1,2,\cdots
\fe
The corresponding scaling dimension of primaries in the deformed CFT $\mathbb{A}_\lambda$ is thus
\ie
\Delta = {1\over 2} \lambda^2 (\ell^2-1) +{\cal O}(\lambda^3).
\fe
This is precisely in agreement with the spectrum of low-lying spectrum of the $A_{k}$ minimal model \cite{DiFrancesco:1997nk} at large $k$, namely the $(r,s)$ degenerate primaries with $r=s=\ell$, whose conformal weights are
\ie
h_{\ell, \ell} = \wt h_{\ell, \ell} = {\ell^2-1\over 4(k+2)(k+3)},
\fe
provided the identification
\ie\label{lambdak}
\lambda = {1\over k} + {\cal O}(k^{-2}).
\fe
The latter is also such that the central charge of $\mathbb{A}_\lambda$ (\ref{cadefa}) agrees with that of the minimal model, namely $c=1- {6\over (k+2)(k+3)}$, up to ${\cal O}(\lambda^3)$ corrections which we have not computed in SFT. Thus, we arrive at the inevitable and yet surprising conclusion that the deformed CFT $\mathbb{A}_\lambda$ is in fact the unitarity $A$-series minimal model!

\section{Discussion}
\label{sec:discuss}

In this paper we have formulated deformations of a 2D CFT in terms of solutions to the classical bosonic closed string field theory (SFT), and a strategy for extracting the deformed CFT data from the spectrum and amplitudes of on-shell fluctuations around the SFT solution. While the SFT formulation at first appears cumbersome due to the freedom in choosing the string vertices and the presence of $bc$ ghosts, these are in fact features that reveal the underlying structure of scheme dependence in Wilsonian renormalized conformal perturbation theory and the important role of the dilaton. The SFT approach is particularly suitable for dealing with noncompact CFTs whose the operator spectra are continuous and there are infinitely many nearly marginal deformations. As illustrated by the example of Horowitz-Polchinski solution, the naive conformal perturbation theory approach would involve marginal deformation operators that are singular in the target space, whereas the SFT deals directly with what amounts to the renormalized coupling or renormalized deformation operators, and produces regular solutions that reduces to those of the spacetime effective field theory in the slow-varying field approximation.

A particularly striking outcome of our analysis is the perturbatively-marginal $\cos(\sqrt{2} X)$ deformation of the noncompact free boson, which results in a condensation of tachyon and dilaton, and the emergence of the ends of space described by Runkel-Watts (RW) theory. While RW theory was originally discovered as the infinite $k$ limit of the $A_{k}$ minimal model, we find that the spectrum of minimal model can be recovered from the standing waves of tachyon between RW walls, up to higher order corrections to the identification of the deformation parameter $\lambda$ with $1/k$. What we have not explained, however, is the quantization of the deformation parameter $\lambda$ intrinsically from the SFT perspective. This requires going beyond the slow-varying field approximation, and possibly understanding the convergence of perturbative closed SFT.

A slightly awkward feature of our SFT construction for CFT deformations that change the central charge is the necessity for introducing an extraneous linear dilaton ``$Y$-sector'', and in principle having to fine tune the SFT solution that maintains the physical decoupling of this sector. An alternative possible approach is to construct an ``off-critical SFT'', based on a worldsheet CFT with nonzero total central charge and BRST anomaly, where the zero string field is not a solution, while actual solutions to the SFT equation would recover the physics of a deformed and critical worldsheet CFT.

Conventional formulation of closed SFT requires rather complicated choice of string vertices that play the role of Feynman vertices and are by construction symmetric with respect to string fields. We have introduced a string field frame in which the string vertices are asymmetric, where one string field is viewed as outgoing and the rest incoming. This is particularly suitable for analyzing solutions to the SFT equation and fluctuations around the solution, as one can work in the ``flat-vertex frame'' in which the string bracket can be computed without applying nontrivial conformal transformations on the string fields. In the end, the string fields in the flat-vertex frame can be related to the string fields in a conventional frame by an in-principle known field redefinition. The technical simplification due to the flat-vertex frame has played an important role in simplify our analysis of SFT equations in the slow-varying field limit. It is of interest to systematize computations in the flat-vertex frame, which may facilitate practical implementation of higher order renormalized conformal perturbation theory and shed light on its convergence property.

Finally, we hope further analysis of string field solutions in the slow-varying field approximation will elucidate the nature of diffeomorphism in SFT \cite{Ghoshal:1991pu}, particularly concerning horizons in the spacetime, and clarify the meaning of the worldsheet CFT for time-dependent string backgrounds.

\section*{Acknowledgements}

We would like thank Bruno Balthazar, Minjae Cho, Carlo Maccaferri, and Ashoke Sen for discussions. This work is completed during the program ``What is String Theory? Weaving Perspectives Together'' at Kavli Institute for Theoretical Physics, Santa Barbara, and we thank KITP for its hospitality and stimulating atmosphere. This work is supported by DOE grant DE-SC0007870.

\appendix

\section{Runkel-Watts CFT data}
\label{sec:runkelwatts}

The Runkel-Watts theory \cite{Runkel:2001ng} is a $c=1$ irrational unitary CFT with primaries $\phi_p$ of conformal weight $h=\wt h = {p^2\over 4}$, where $p\in \mathbb{R}_+ \backslash \mathbb{Z}_+$. For convenience we will extend the parameter $p$ to negative real values by defining $\phi_{-p} = \phi_p$. We further define $\phi'_p \equiv (-)^{\lfloor p \rfloor}\phi_p$, with $\phi'_{-p} = -\phi'(p)$. As such the two-point functions are normalized according to
\ie
\left\langle \phi'_{p_1}(z_1) \phi'_{p_2}(0)\right \rangle = \left(\delta(p_1-p_2) - \delta(p_1+p_2) \right) |z|^{-p_1^2},
\fe
and the three-point functions take the form
\ie
\left\langle \phi'_{p_1}(z_1) \phi'_{p_2}(z_2) \phi'_{p_3}(z_3)\right \rangle = {c'(p_1, p_2, p_3) \over |z_{12}|^{p_1^2+p_2^2-p_3^2\over 2}  |z_{23}|^{p_2^2+p_3^2-p_1^2\over 2}  |z_{13}|^{p_1^2+p_3^2-p_2^2\over 2}}.
\fe
The structure constant $c'(p_1, p_2, p_3)$ is given by
\ie
c'(p_1, p_2, p_3) = P'(p_1, p_2, p_3) e^{Q(p_1, p_2, p_3)},
\fe
where $Q$ is given in equation (10) of \cite{Runkel:2001ng}, and $P'(p_1, p_2, p_3)$ is related to the function $P(p_1, p_2, p_3)$ defined in equation (9) of \cite{Runkel:2001ng} by the sign $(-)^{\lfloor p_1 \rfloor+\lfloor p_2 \rfloor+\lfloor p_3 \rfloor}$ and the aforementioned extension to negative values of $p_i$. The sign change in our choice of basis is such that $P''(p_1, p_2, p_3)$ can be expressed as
\ie
P'(p_1, p_2, p_3) &= {1\over 2} \big(\Theta(p_1 + p_2 - p_3) + \Theta(p_1 + p_3 - p_2) + \Theta(p_2 + p_3 - p_1) + \Theta(-p_1 - p_2 - p_3) - 2 \big)
\\
&~~~ - {1\over 2} \sum_{\epsilon_i = \pm} \sum_{n=1}^\infty \epsilon_1\epsilon_2 \epsilon_3 \Theta(\epsilon_1 p_1 + \epsilon_2 p_2 + \epsilon_3 p_3 -2n),
\fe
where $\Theta$ is the Heaviside step function.

The function $Q$ is defined as the analytic continuation of 
\ie\label{qqdef}
	Q(p_1, p_2, p_3) = \int_0^1 \frac{d \beta}{(- \log \beta) (1 - \beta)^2} R(p_1, p_2, p_3, \beta) 
\fe
from the domain $0<p_i<1$, $p_1+p_2+p_3<2$, where $R$ is defined as
\ie
	R(p_1, p_2, p_3, \beta) = 2 + \sum_{\epsilon=\pm}\sum_{i=1}^3 \beta^{\epsilon p_i} - \sum_{\epsilon_i = \pm} \beta^{{1\over 2} \left(\epsilon_1 p_1 + \epsilon_2 p_2 + \epsilon_3 p_3\right)} .
\fe
Note that near $\beta=1$, $R$ has a fourth order zero whereas the denominator in the integrand of (\ref{qqdef}) has a third order zero, and the integral is convergent. To analyze the properties of $Q$, it useful to write the RHS of (\ref{qqdef}) as
\ie
\int_0^1 d\beta \int_0^\infty du \sum_{n = 1}^{\infty} n \beta^{u + n - 1} R(p_1, p_2, p_3, \beta).
\fe
The absolute convergence of the summation and integration allows for swapping their order, giving
\ie
	Q(p_1, p_2, p_3 ) &= \sum_{n = 1}^\infty n \int_0^\infty du \int_0^1 d \beta \beta^{u + n - 1}  R(p_1, p_2, p_3, \beta) \\
	&= \sum_{n = 1}^\infty n \int_0^\infty du \left(\frac{2}{u + n} + \sum_{\epsilon = \pm}\sum_{i=1}^3 \frac{1}{u + n + \epsilon p_i} - \sum_{\epsilon_i = \pm} \frac{1}{u + n + {1\over 2} \sum_{i=1}^3 \epsilon_i p_i }\right) \\
	&= \sum_{n = 1}^\infty n \left(-2 \log n - \sum_{\epsilon = \pm}\sum_{i=1}^3 \log (n + \epsilon p_i) + \sum_{\epsilon_i = \pm} \log \big(n + {1\over 2} \sum_{i=1}^3 \epsilon_i p_i  \big) \right).
\fe
In particular, note that in the limit of simultaneously scaling $p_1, p_2, p_3$ to 0, $Q$ vanishes at quartic order. 
%Further inspecting
%\begin{equation}
%	e^{Q(p_1, p_2, p_3)} = \prod_{n=1}^\infty \prod_{\epsilon = \pm} \prod_{i=1}^3 \left(\frac{n e^{\frac{\epsilon v}{n} - \frac{v^2}{2 n^2}}}{n + \epsilon v}\right)^n \prod_{n>0, \epsilon_x, \epsilon_y, \epsilon_z = \pm} \left(\frac{2n + \epsilon_x x + \epsilon_y y + \epsilon_z z}{2n e^{\frac{\epsilon_x x + \epsilon_y y + \epsilon_z z}{2n} - \frac{(\epsilon_x x + \epsilon_y y + \epsilon_z z)^2}{8n^2}}}\right)^n 
%\end{equation}
We further observe that $e^Q$ has a pole of order $|n|$ when any one of $p_1, p_2, p_3$ takes an integer value $n$, and a zero of order $n$ when $\sum_{i=1}^3 \epsilon_i p_i = 2n$ for some positive integer $n$ and $\epsilon_i=\pm$. 
One can verify that when $P'(p_1, p_2, p_3)$ is nonzero, $e^{Q(p_1, p_2, p_3)}$ remains finite. %This requires a bit of casework on absolute values of the integer parts when counting poles at the corners of the tetrahedra. 

Now we pass to the target position space by Fourier transforming the basis of primaries $\phi_p'$ with respect to $p$, and consider %\footnote{{\it ``Only perverts think in momentum space.” - Lenny Susskind.}}
\ie
\wt \phi_x \equiv \int dp\, e^{i p x} \phi'_p.
\fe
While $\wt\phi_x$ does not have a definite conformal weight, we are particularly interested in the large $x$ limit where the small weight component dominates and the coordinate dependence of correlation functions can be ignored, so that the three-point function of $\wt \phi_x$ can be expressed as
\ie
\wt c (x_1, x_2, x_3) \approx \int {dp_1 dp_2 dp_3} e^{i p_1 x_1 + i p_2 x_2 + i p_3 x_3} c'(p_1, p_2, p_3).
\fe
Furthermore, the large distance behavior of $\wt c$ is controlled by the momentum space discontinuity of $c'$, particularly along the $\pm p_1 \pm p_2 \pm p_3 = 0$ planes. Near $p_1 + p_2 + p_3 = 0$, for instance, where $Q(p_1, p_2, p_3)$ vanishes, the discontinuity in $c'(p_1, p_2, p_3)$ is captured by
\ie
 - \frac{1}{2} \Theta(p_1 + p_2 + p_3) .
\fe
Taking its Fourier transform gives the contribution to $\wt c (x_1, x_2, x_3)$,
\ie
	 -\frac{1}{2} \int {dp_1 dp_2 dp_3} \left(\frac{1}{i x_1} \partial_{p_1} e^{i p_1 x_1 + i p_2 x_2 + i p_3 x_3}\right) \Theta(p_1 + p_2 + p_3) 
%	&= -\frac{i}{2x_1} \int dx dy dz e^{i q_1 x + i q_2 y + i q_3 z} \delta(x + y + z) \\
	= -\frac{i}{2x_1} (2 \pi)^2 \delta(x_{12}) \delta(x_{13}).
\fe

\bibliographystyle{JHEP}
\bibliography{CPT}

\end{document}